\documentclass{aa}  

\usepackage{graphicx}
\usepackage{txfonts}

\newcommand{\te}{TESS}
\newcommand{\kms}{km\,s$^{-1}$}

\begin{document}

   \title{\te\ observations of non-Be fast rotators}

   \author{Ya\"el Naz\'e
          \inst{1}\fnmsep\thanks{F.R.S.-FNRS Senior Research Associate}
          \and
          Nikolay Britavskiy\inst{1,2}
          \and
          Jonathan Labadie-Bartz\inst{3}
          }

   \institute{Groupe d'Astrophysique des Hautes Energies, STAR, Universit\'e de Li\`ege, Quartier Agora (B5c, Institut d'Astrophysique et de G\'eophysique), \\
All\'ee du 6 Ao\^ut 19c, B-4000 Sart Tilman, Li\`ege, Belgium\\
              \email{ynaze@uliege.be}
         \and
             Royal Observatory of Belgium, Avenue Circulaire/Ringlaan 3, 1180 Brussels, Belgium
         \and
         LESIA, Paris Observatory, PSL University, CNRS, Sorbonne University, Universit\'e Paris Cit\'e, 5 place Jules Janssen, 92195 Meudon, France
            }


 
  \abstract
   {}
   {The variability of fast-rotating Oe/Be stars has been reported in detail in recent years. However, much less known about the behaviour of fast-rotating OB stars without known decretion disks, and hence it is difficult to identify the commonalities and differences in the photometric variability of these two populations, especially with regards to their pulsational properties and their link with the presence of circumstellar material.}
   {Via an in-depth literature search, we identified a set of fast-rotating ($v \sin(i)>200$\,\kms) early B-type stars not known to have disks. \te\ and {\it Kepler} light curves were built for 58 stars that appear isolated (no bright neighbour within 1\arcmin\ and no known companion) to avoid contamination of the light curves. Frequency spectra were calculated and then analysed to determine the noise level and the presence of significant signals above the noise. }
   {Red noise is detected in all targets, without obvious correlations between noise and stellar parameters. Long-term changes are much less frequent than in Be stars, with only 12\% of our targets having the variability below 0.5\,d$^{-1}$ dominating their frequency spectrum. In contrast, strong frequency groups are detected in about a third of targets, as in Be stars. These groups generally occur in pairs with harmonic frequencies, as is usually seen in Be stars, but with the first group more often displaying larger amplitudes. Finally, the most frequent variability is due to isolated frequencies in the 0.5--6.\,d$^{-1}$ range (which is found in two-thirds of cases and dominates the spectra in 42\% of the sample). Higher-frequency signals (up to 40\,d$^{-1}$) are sometimes also detected but rarely (only 12\% of stars) appear as the strongest ones of the frequency spectra. Overall, fast-rotating B-type stars, with or without disks, display similar photometric properties, except as regards their longer-term behaviour. }
   {}


   \maketitle
%

\section{Introduction}

For more than a century, a subgroup of massive OB stars were found to display, at least from time to time, Balmer emission lines in their spectra. This led to the creation of the ``classical Oe/Be stars'' category. These stars are now understood to be surrounded by a decretion disk, which is responsible for the remarkable emissions (for a review, see \citealt{riv13}). Another particularity shared by objects in this category is fast rotation (see \citealt{fre05} and, for an in-depth review, see \citealt{zor23}). Combined with the detection of low-mass hot companions (e.g. \citealt{wan21}) or of a truncation of disks \citep{kle19}, this suggests that Be stars actually are, at least sometimes, binary interaction products, in line with theoretical proposals \citep{van97,sha14}. The initially more massive star of the system has lost mass which was at least partially accreted by the companion. Angular momentum was also transferred during the event, speeding up the accreting companion while increasing its mass and it eventually became an Oe/Be star. 

When monitored, such stars display changes on various timescales. First, there are ejections of the material building the disk, which then clears from the inside out. This should correspond to brightenings or dimmings, depending on geometry (see the numerous observational and theoretical studies, e.g. \citealt{hub98,kel02,men02,sig13,gor21} and references therein), followed by a return to the baseline occurring on days, weeks, or months. Second, there are non-radial pulsations, often at low frequencies ($<$5\,d$^{-1}$, \citealt{cuy89}) and/or rotational signals \citep{bal21}. Finally, there is also stochastic red noise, as usual in massive stars \citep{naz20,lab22}.

In recent years, the advent of space facilities, such as {\it CoRoT} (Convection, Rotation and planetary Transits), {\it Kepler}, or \te\ (Transiting Exoplanet Survey Satellite), as well as long-term campaigns by ground observatories (e.g. {\it KELT}), has enabled astronomers to examine the variability of Oe/Be stars in detail \citep{gut07,lab17,see18,bal20,naz20,lab22}. Statistics were established on the occurence rates of the different variation types, and their typical characteristics were derived. However, the impact of the presence of disk circumstellar material on this variability on timescales of a few days and shorter has not yet been fully clarified.

In this study, we decided to use high-cadence photometric data from the space telescopes \te\ and {\it Kepler} to offer a counterpoint to the Be star studies. Observational studies of non-Be fast-rotating massive stars also suggest that such stars arise from binary interactions (e.g. O stars in \citealt{bri23}), which would mean that they share a common origin with Oe/Be stars. However, if they do not belong to the Oe/Be category, then circumstellar material is absent. The comparison of the two samples - fast-rotating Be and non-Be stars (sometimes called `Bn' stars) - would then clarify the impact of disks on the recorded variability, and/or any intrinsic differences in photospheric signals. The study of non-Be stars could also help us understand the origin of stellar rotation. Overall, fast rotation can be primordial or arise from a post-mass transfer spin-up or even a merger event. Any difference in the observed variability could be linked to differences in the inner layers. It is therefore important to have a significant sample of well-studied targets from both Be and non-Be domains in order to systematise their properties and draw a global picture. The paper is divided as follows: Sect. 2 presents the sample and the datasets, while Sect. 3 focuses on red noise and Sect. 4 on the other variabilities. A conclusion then summarises the results.

\begin{figure}
  \centering
  \includegraphics[width=9cm]{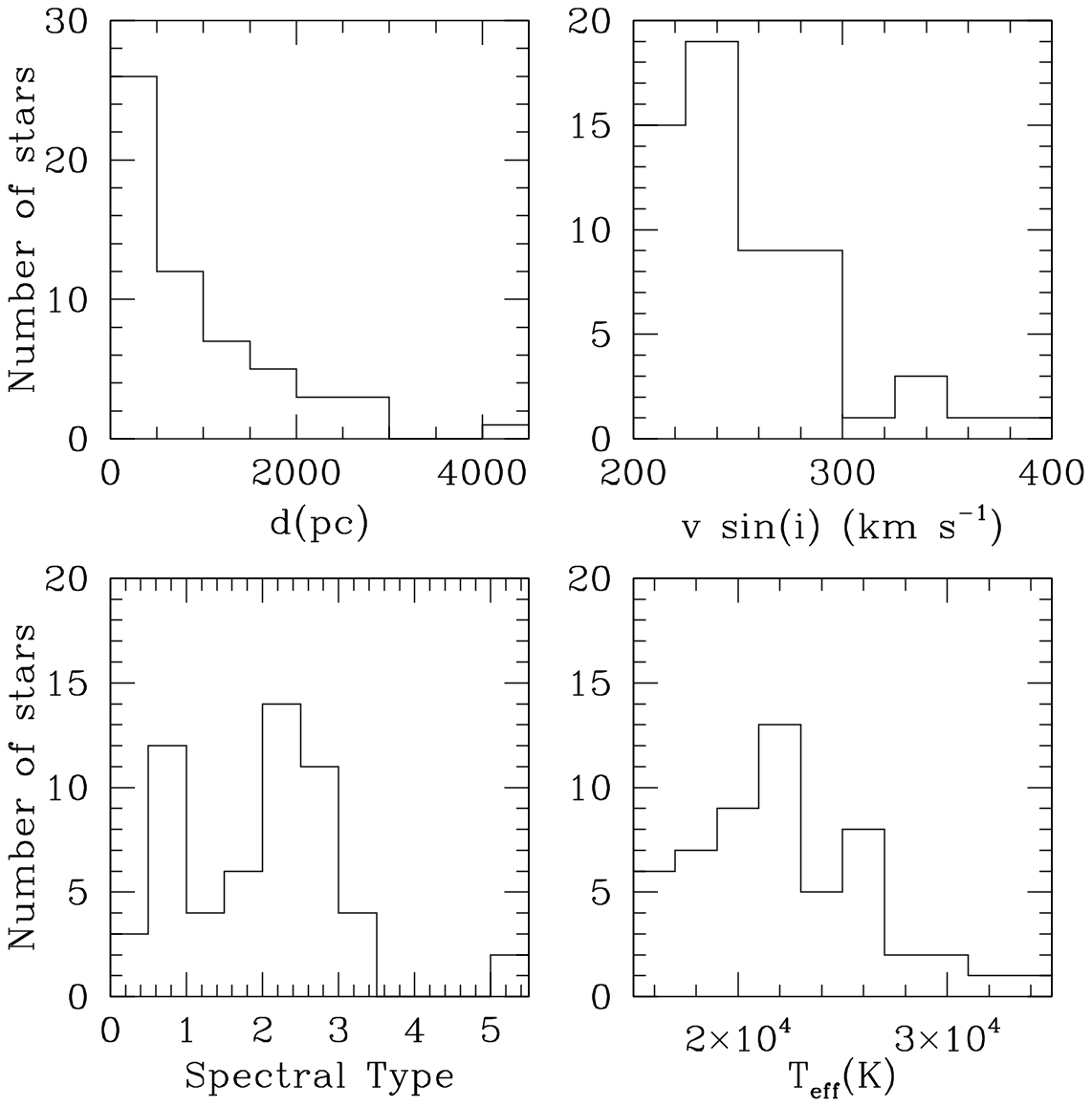}
  \caption{Histograms of the distances, projected rotational velocities, spectral types, and temperatures of our targets. 
  }
  \label{histo}
\end{figure}

\section{Observing fast rotators}
\subsection{Defining the sample}
While several studies have investigated the rotation of massive stars, often for a specific cluster, a single overall catalogue of precise projected rotational velocities of stars remains difficult to find. Therefore, we decided to build our own sample. To this aim, we performed an in-depth search of the literature in several steps. First, the Vizier database\footnote{https://vizier.cds.unistra.fr/} was explored to identify catalogues dedicated to rotation rates. This led to the identification of a large series of papers \citep{abt02,bra12,coc20,daf07,dav15,gar15,2005csss...13..571G,han18,how97,hua06,hua10,lev06,mar12,pen09,sim14,str05,van12,wol07,xia22}. Their associated tables were then filtered to keep only stars with fast rotation, defined as $v \sin(i)>200$\,\kms. This choice of threshold was done because, above it, most objects should have been spun up by binary interactions \citep{dem13}. In addition, we kept only the hottest B-type main sequence stars, that is those with spectral types between B0V and B3V (with luminosity class IV also allowed, but not supergiants as such evolved stars do not display Be features) or with effective temperatures $T_{eff}$ in 17000--31500\,K and $\log(g)>3.7$ (the chosen criterion depends on the catalogue as some catalogues uses spectral types, while others rely on temperature and gravity). 

Second, we eliminated from this list known Be stars (either listed as such in SIMBAD\footnote{http://simbad.u-strasbg.fr/simbad/} or belonging to the BeSS catalogue\footnote{http://basebe.obspm.fr/basebe/ and \citep{nei11}, from Fall 2022}) as we wish to investigate fast rotators but {\it without} disks. We also discarded obvious binaries (``SIMBAD categories EB*, SB*, and El*''), magnetic stars, and young objects (``SIMBAD category YSO*'') to avoid contamination by unrelated objects. Finally, to be certain to get a light curve from the fast-rotating object despite the rather coarse spatial resolution of \te, we also took out stars possessing any nearby bright companions in the {\it GAIA} Data Release 3 catalogue (a companion within 1\arcmin\ of the star and with $\Delta(G)<2.5$\,mag).

As a last step, literature was examined in detail for each object of this cleaned list. This allowed us to discard further binaries as well as objects whose stellar parameters in recent studies lie outside of our selection criteria. We note that a few objects listed as B5 but with temperatures in our allowed temperature range were kept, as were a few cases of slightly too hot or too cold stars listed with allowed spectral types. Our final list comprises 61 objects, among which are several well-known pulsators. Five of them were not observed by \te, but two of these five have {\it Kepler} (K2) light curves. Our sample therefore comprises 58 stars in total (see Table \ref{list} for details).

Distribution of distances (from \citealt{bai21}), projected rotational velocities, spectral types, and temperatures are provided in Figure \ref{histo}. These histograms show that two-thirds of the targets lie closer than 1\,kpc, and 90\% of them display projected rotational velocities between 200 and 300\,\kms. Targets are also nearly evenly distributed between spectral types B0--1.5 and B2--3.

Using {\sc stilism}\footnote{https://stilism.obspm.fr/} \citep{lal14}, we calculated the reddenings based on the SIMBAD galactic coordinates and the known distances. Bolometric luminosities were then obtained in the usual way using the observed $V$ magnitudes (from Simbad), the derived reddenings (assuming $R_V=3.1$), the known distances, and bolometric corrections $BC_V$ calculated from the effective temperatures using the formula of \citet{ped20}. In four cases, temperatures were not known; hence, typical temperatures and bolometric corrections for the targets' spectral types were taken from \citet{mam13}\footnote{see {\scriptsize http://www.pas.rochester.edu/$\sim$emamajek/EEM\_dwarf\_UBVIJHK\_colors\_Teff.txt}}. In four other cases, there is no $V$ magnitude in SIMBAD, no distance in \citet{bai21}, or no reddenings (distances are larger than the maximum distance allowed by {\sc stilism}) hence typical bolometric luminosities for the targets' spectral types were used. These cases are identified by $^m$ in Table \ref{list}. To conclude on this subject, one remark must be made. Most of our targets were rarely studied and their parameters often come from global studies, not papers dedicated to one specific target. The stellar parameters may thus suffer from some uncertainties, especially since their derivation is quite delicate for fast-rotating stars (see e.g. \citealt{fre05}). As a check, we compared the derived bolometric luminosities to those expected in Mamajek's calibration for the targets' spectral types (or temperatures, for the two B5 cases and the two cases without known spectral types): two-thirds of the targets had these luminosities agreeing within a factor of two (or 0.3\,dex, i.e. about half a subtype) and only 13\% had them differing by more than a factor of four (or 0.6\,dex). As a second check, we compared the targets' temperatures to the expected temperatures for their spectral types: the difference was less than 2\,kK for 52\% of stars and more than 5\,kK for only 15\% of them. For CPD-50\,9216 and CPD-48\,8710, the observed mismatches could be easily solved as their temperatures and derived luminosities appear fully compatible with a B2 spectral type: this suggests a (small) misclassification in their cases. Overall, larger differences are found for more distant ($>1$\,kpc) targets. This is unsurprising as distant objects are certainly amongst the least studied objects of our sample. Nevertheless, considering the difficulty to secure parameters of fast rotators, the large fraction of good agreements may be underlined.

As a last step, we estimated $v\sin(i)/V_c$. In view of the uncertainties in the stellar parameters (see above), no attempt was made to correct for inclination. For the same reason, we cannot securely match the targets to the closest evolutionary tracks to derive ages or critical velocities. A crude estimate was then calculated from $V_c=\sqrt{GM_*/R_*}$ with the stellar radius evaluated from $L_{\rm BOL}=4\pi R_*^2\sigma T^4_{eff}$ and the stellar masses $M_*$ the typical ones for the considered spectral types in Mamajek's calibration. Table \ref{list} provides the $v\sin(i)/V_c$ (see Section 4 for further details), except for two targets without spectral types and two others with a (most probably incorrect) B5V type. Our targets lie mostly between ratios of 0.3 and 0.6. No target is detected with a very low ratio but this is a known observational bias: while fast-rotating Be stars can be recognised at nearly all inclinations as their detection is based on the presence of H$\alpha$ emission, fast-rotating B-stars that are not surrounded by circumstellar material can only be recognised from their broad lines, in other words if they are not seen too close to pole-on. In the Bn sample of \citet{coc20}, which is mostly composed of stars with later spectral types, the lowest $v\sin(i)/V_c$ ratios were around 0.4, the highest ones near 1.0, and the average around 0.6: their stars thus display, on average, somewhat higher ratios than stars in our sample. The ratios observed for our stars also peak at slightly lower values than for the Be sample of \citet{coc20}. We also calculated critical velocities from {\it GAIA} masses and radii \citep{ker22}. This resulted in an average increase in the $v\sin(i)/V_c$ ratios of 50\%, with most values in 0.5--0.7 and a few ratios now larger than 1. However, such {\it GAIA} parameters are known to be quite uncertain for the hottest stars. In the literature, critical velocities have been published for three stars (HD\,5882, HD\,35532, and HD\,87015; see \citealt{hua10}) and these values are 30\% lower than our first estimates for these stars and 10\% than the second ones. Overall, our first estimates may thus be biased towards slightly too low ratios (and the second estimates towards too large ones). Nevertheless, we kept these first estimates for the moment, to be conservative (and avoid biasing towards too large values) while awaiting for future, better determinations from detailed spectroscopic analyses.

\subsection{Light curves}
Launched in April 2018, the \te\ satellite \citep{ric15} provides photometric measurements for nearly all of the celestial sphere. Observations are organised by ``sectors'' with durations of about 25\,d, and this paper uses data up to Sector 74. All but one of our \te\ targets have been observed in several Sectors (generally 2--5, but up to 19); these light curves were examined independently. In each sector, preselected stars are measured every 2\,min, while usual sky images are taken with a cadence of 30\,min up to Sector 26, 10\,min for Sectors 27--55, and 200\,s afterwards. The sector duration leads to a natural peak width in frequency space of 1/25\,d, i.e. 0.04\,d$^{-1}$, while the observing cadences of 30\,min, 10\,min, 200\,s, and 2\,min lead to Nyquist frequencies of 24, 72, 216, and 360\,d$^{-1}$, respectively. The main steps of data reduction (pixel-level calibration, background subtraction, flat-fielding, and bias subtraction) are done by a pipeline similar to that designed for the {\it Kepler} mission.

For 2\,min cadence data, simple aperture photometry (SAP) as well as time series corrected for crowding, the limited size of the aperture, and instrumental systematics (Pre-search Data Conditioning SAP flux - PDC) were downloaded from the Mikulski Archive for Space Telescopes (MAST)\footnote{https://mast.stsci.edu/portal/Mashup/Clients/Mast/Portal.html}. We kept only the best quality (quality flag=0) data. In each case, we compared the SAP and PDC light curves. Generally, both curves were similar or PDC light curves were better. Only in a few cases were SAP light curves preferred as PDC photometry appeared noisier.

For the data with other cadences, individual light curves were extracted for each target from \te\ full frame images. Aperture photometry was done on image cutouts of 50$\times$50 pixels using the Python package Lightkurve\footnote{https://docs.lightkurve.org/}. A source mask was defined from pixels above a given flux threshold (generally 10 Median Absolute Deviation over the median flux, but it was decreased for faint sources or increased if neighbours existed). The background mask was defined by pixels with fluxes below the median flux (i.e. below the null threshold), thereby avoiding nearby field sources. To estimate the background, two methods were used: first, a principal component analysis, usually with five components; second, a simple median. In each case, the two background-subtracted light curves were compared. Generally, principal-component-analysis-corrected light curves were better but in a few cases they appeared noisier hence the curves using the median correction were preferred. 

\begin{figure}
  \centering
  \includegraphics[width=9cm]{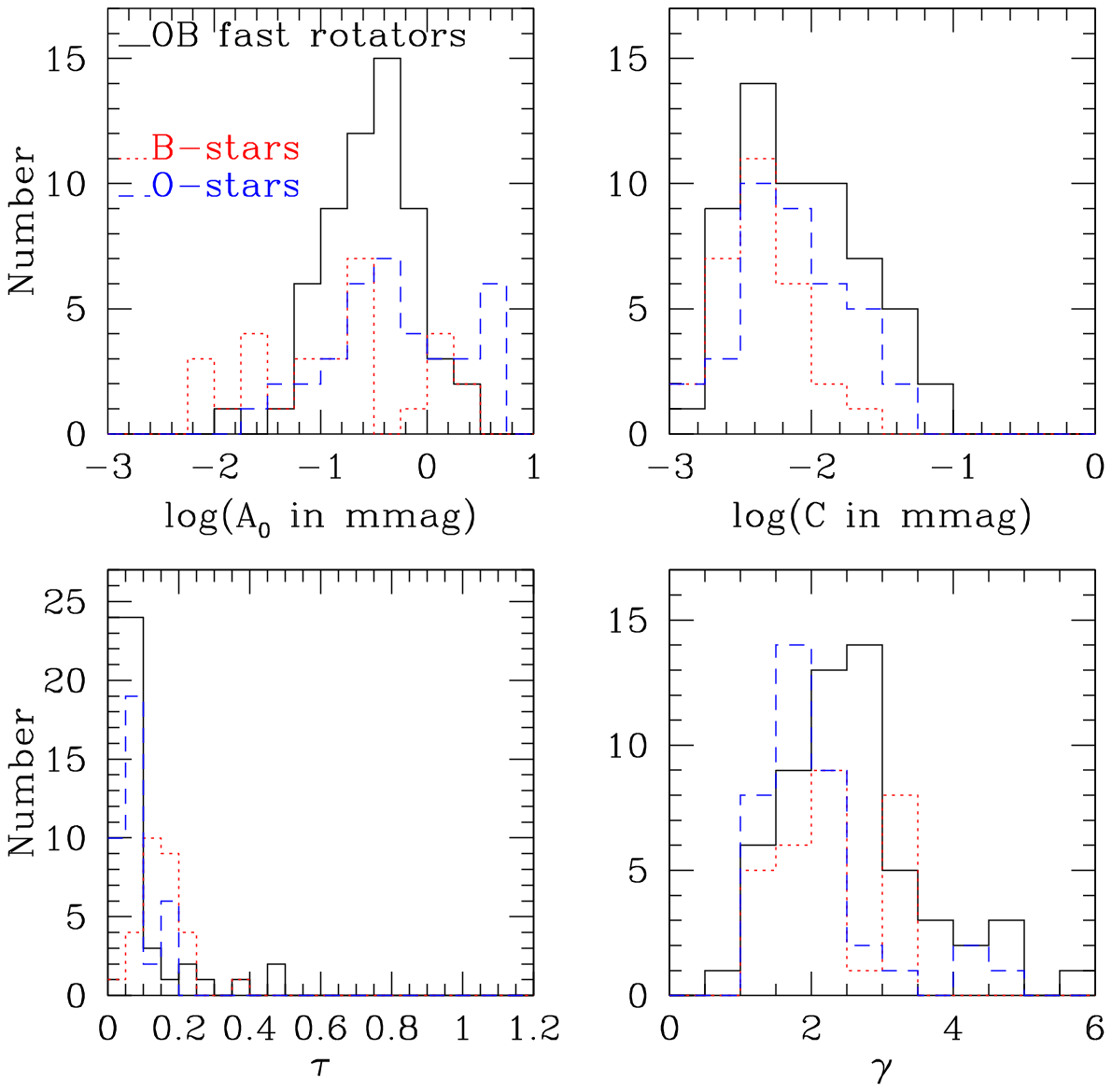}
  \caption{Histograms of red noise parameters for our fast rotators (solid black line), compared to those determined for O and B-type stars by \citet[- dotted blue and red lines, respectively]{bow20}. }
  \label{historn}
\end{figure}

The two targets observed with {\it Kepler} (HD\,138485 and HD\,142378) are bright and have saturated the detector. In such cases, the star spot is larger due to spilled charges. For the latter star, the pipeline-selected aperture was sufficiently large to capture most light hence the pipeline light curve is usable\footnote{With identifier ktwo205104403-c02\_llc, it can be downloaded from MAST.}. This was not the case for the former star, leading to weird results with erratic photometry. Fortunately, \citet{pop19} produced light curves for several saturated stars in the framework of the HALO survey\footnote{https://archive.stsci.edu/hlsp/halo}, among which HD\,138485 (EPIC 200194914). In both cases, as for the 2\,min cadence data, only the best quality data were considered and the corrected curves were favoured over the SAP data. The two targets were observed every 30\,min for about 80\,d with {\it Kepler}, leading to a Nyquist frequency of 24\,d$^{-1}$ and a natural peak width of 0.01\,d$^{-1}$.

For all 221 light curves, the raw fluxes were converted into magnitudes using the usual formula $mag=-2.5\times \log(flux)$ and the mean was then subtracted. For the \te\ data with cadences larger than 2\,min, data points with errors larger than the mean of the errors plus three times their 1$\sigma$ dispersion were also discarded. In addition, a few isolated outliers and a few short temporal windows with sudden high scatter were eliminated from the few affected light curves. For seven targets, a trend was detected in one sector but not others, suggesting an instrumental effect: the linear or parabolic best-fit of the trend was thus calculated and taken out. Finally, the frequency spectrum of each light curve was calculated using a modified Fourier algorithm taking into account the presence of gaps \citep{hmm,gos01,zec09}. The frequency resolution was set to one tenth of the natural peak width\footnote{We also consider this value as representative of the frequency uncertainty, although some authors are even more conservative (e.g. \citealt{kal08} favours a value of one fourth of the peak width). }. Figures showing all light curves and their associated frequency spectra are available on Zenodo\footnote{https://zenodo.org/doi/10.5281/zenodo.12699682}. 

\begin{figure*}
  \centering
  \includegraphics[width=7.2cm]{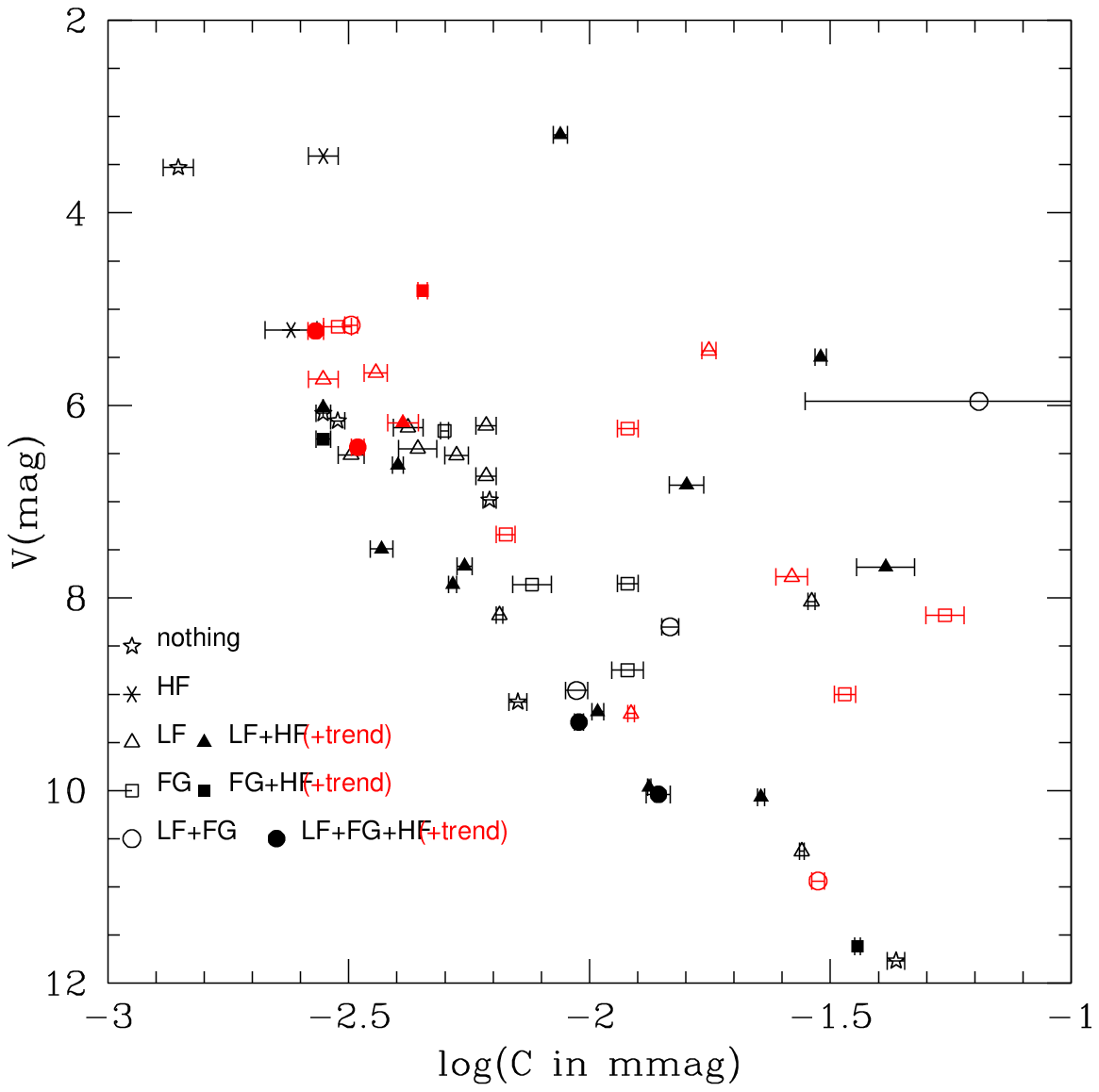}
  \includegraphics[width=7.2cm]{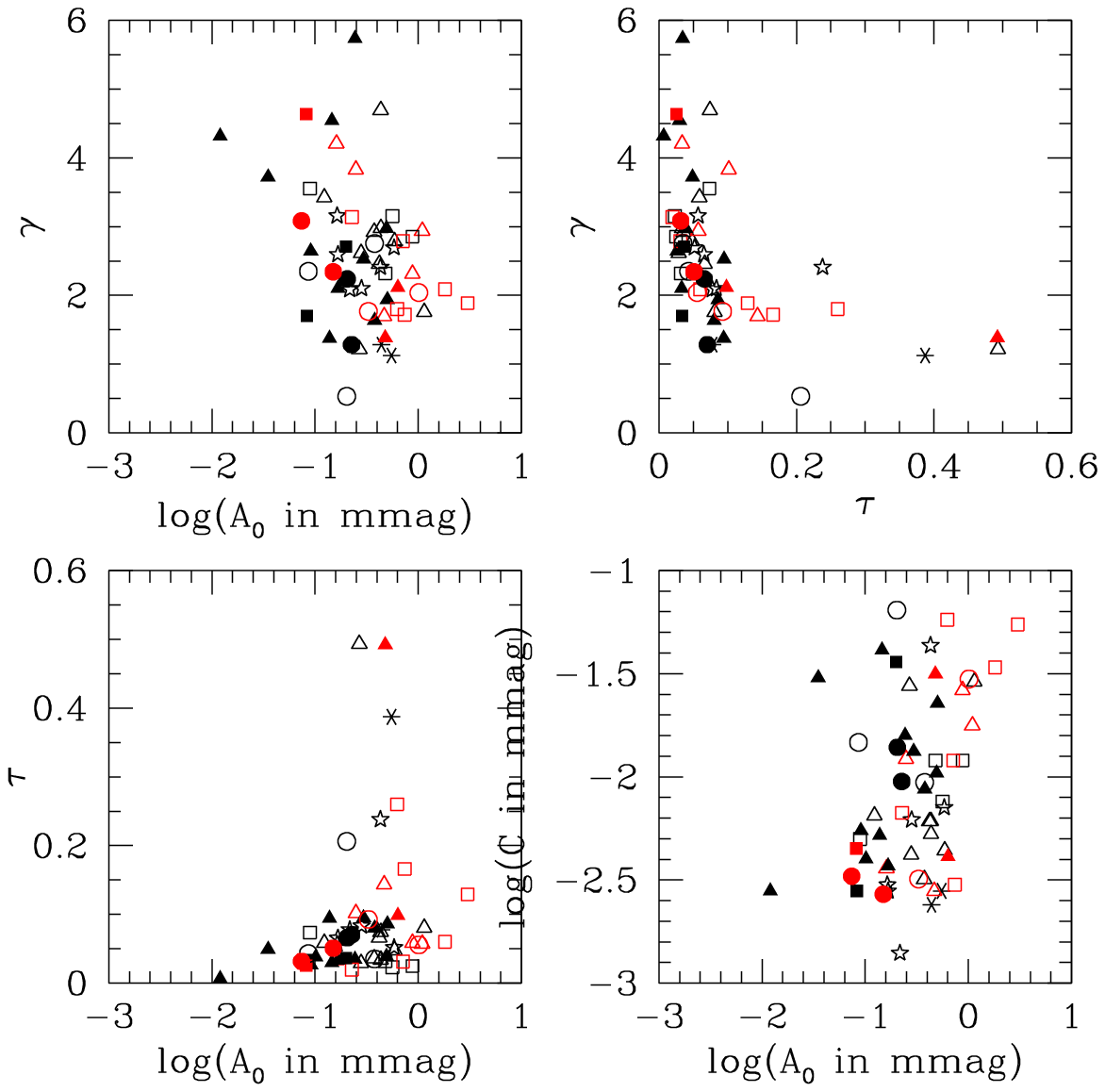}
  \includegraphics[width=3.6cm]{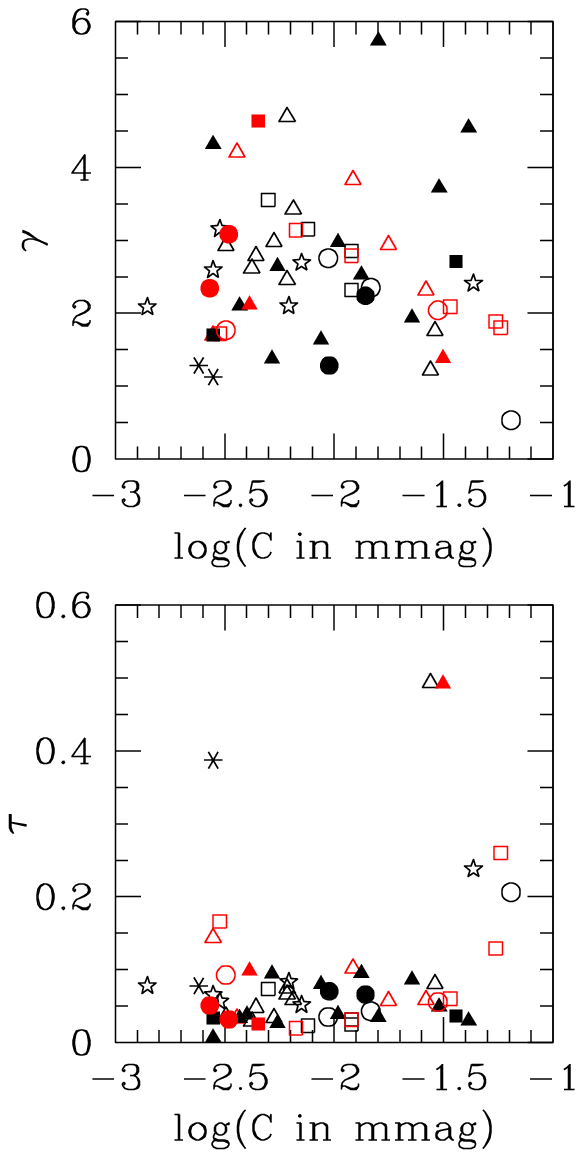}
  \caption{Comparison between red noise parameters. {\it Left:} Comparison between the white noise level, $C$, and the $V$ magnitude of all targets. Symbols indicate the type of recorded variability (see Sect. 4): objects without any variability beyond red noise are shown as stars, objects with only high-frequency (HF) signals as asterisks, objects with dominating low-frequency (LF) signals as triangles, objects with strong frequency groups (FG) as squares, and objects with both LF and FG signals as circles. For the last three categories, the symbols are filled if high-frequency signals also exist, and empty otherwise. If trends or very low-frequency signals are detected, the symbols are shown in red. {\it Right:} Two-by-two comparison between red noise parameters, using the same symbols.}
  \label{cmag}
\end{figure*}

\section{Red noise}
With the advent of space-based high-cadence photometry, it became clear that massive stars (with types O, B, WR, or LBV) generally display a gradual increase in power towards low frequencies, i.e. a so-called `red noise' \citep{blo11,rau19,bow19,naz21}. Our sample is no exception: all targets display red noise. Several origins have been proposed to explain this ubiquitous variability, most notably a combination of internal gravity waves excited at the interface between the convective core and the radiative envelope \citep{Rog13} or in a subsurface convection zone \citep{blo11}. 

Characterising the properties of this noise is important for two reasons. First, its characteristics should be linked to the physical processes giving rise to it; hence, they could help us better understand its origin. Second, white+red noise represents a varying background to be taken into account while assessing the significance of any other signal and hence needs to be known. The variation with frequency of the white+red noise is usually fitted using the generalised formalism of a suggestion by \citet{har85}:
\begin{equation}
  A(\nu) = C+ \frac{A_0}{1 + (2\,\pi\,\tau\,\nu)^{\gamma}}
  \label{eqrednoise}
\end{equation}
where $A(\nu)$ is the amplitude of the frequency spectrum at frequency $\nu$\footnote{Formally, stochastic processes such as white and red noises sum up quadratically and not linearly; hence, power should be fitted by this formula rather than amplitude. Custom in the massive star field however favours using the latter, which we followed. We note that the results, such as the detection of physical trends, are not affected by this choice (see \citealt{naz21} for a detailed comparison).}, $A_0$ is the red noise level at null frequency, $\tau$ the mean lifetime of the structures producing the red noise, $\gamma$ the slope of the linear decrease, and $C$ the white noise level. Before the fitting, done through a Levenberg-Marquardt algorithm, the larger peaks of the frequency spectra were excised using a simple amplitude threshold to avoid biasing the red noise background estimate. 

As there is no formal error on frequency spectrum amplitudes, the errors on the red noise parameters are difficult to formally calculate. As a first approximation, we followed the procedure outlined in \citet{naz21}: parameter errors were assumed to be equal to the square root of the diagonal elements of the best-fit variance-covariance matrix, multiplied by the square root of $\chi^2({\rm best\,fit})/(0.5 \times N_{\rm data}-4)$ . This assumes that the best-fit reduced $\chi^2$ should amount to one and that the actual number of independent frequencies is half the number of data points in the light curve. 

Appendix B provides the best-fit noise parameters for the 221 frequency spectra. It should be noted that the white noise level is usually reached only at very high frequencies, that is frequencies usually much larger than the Nyquist frequency of 24\,d$^{-1}$ of the lowest cadence data (identified by {\it Kepler} and \te\ Sectors 1--26 without asterisk in Col. 2 of Tables \ref{list} and \ref{rednt} in Appendix). Therefore, the parameters associated with these lower cadence light curves must be taken with caution, whereas those taken with the two highest cadence (i.e. \te\ Sectors $>$55 or sectors with an asterisk in the tables). We note also that the errors may still be somewhat underestimated, as similar light curves may have parameters varying by more than their 1$\sigma$ errors.

To perform comparisons, a single set of parameters should be used for each star. We therefore choose the values associated with the light curve with the highest cadence. If several such curves are available for a given target, the values from the most recent observation are usually chosen (parameters generally appear similar except in a few cases for which that best-fit curve visually seems obviously less suitable than that of another sector). The parameters' distribution is compared to that of other massive stars in Fig. \ref{historn}. As can be seen, the parameter values found for our fast rotators are in excellent agreement with the ranges found for a set of 37 O stars and 29 early B-type stars by \citet{bow20}. Fast rotation thus does not affect much the physical processes responsible for these noises.

We compare the white noise levels $C$ to the targets' magnitudes in Fig. \ref{cmag}. As was found for evolved massive stars \citep{naz21}, the range of white noise levels appears restricted for the faintest stars; in other words, lower $C$ values can only be found for the visually brighter objects. This could be expected as it simply reflects a larger impact of photon noise on the recorded light curves from faint stars: for them, a low intrinsic noise level may remain hidden. As this instrumental effect is not frequency-dependent, it does not affect much the red noise level $A_0$. In this context, it could be noted that the red noise level always appears larger than the white noise level, with an average ratio of $\sim$50 (Fig. \ref{historn}).

\begin{figure}
  \centering
  \includegraphics[width=9cm]{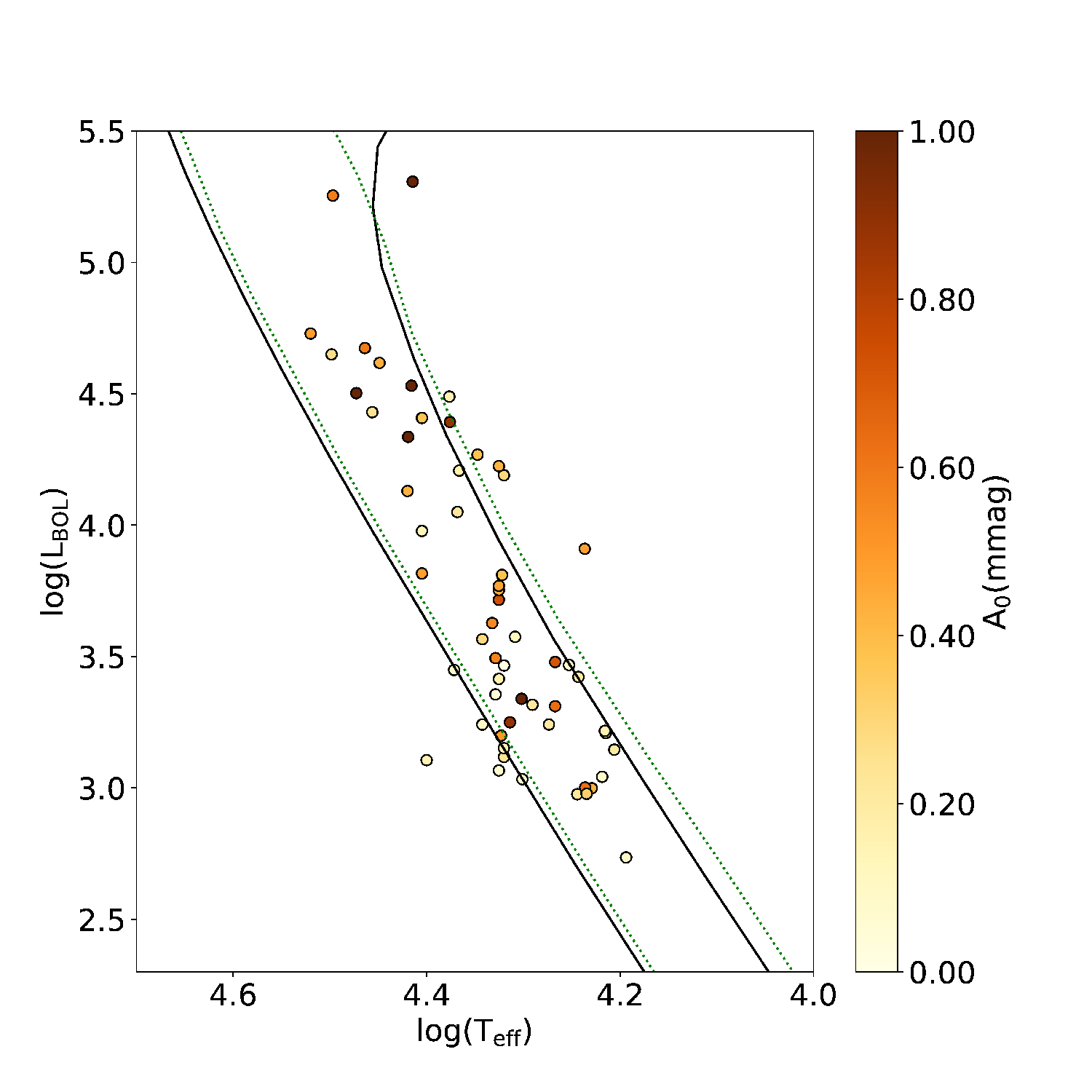}
  \includegraphics[width=9cm]{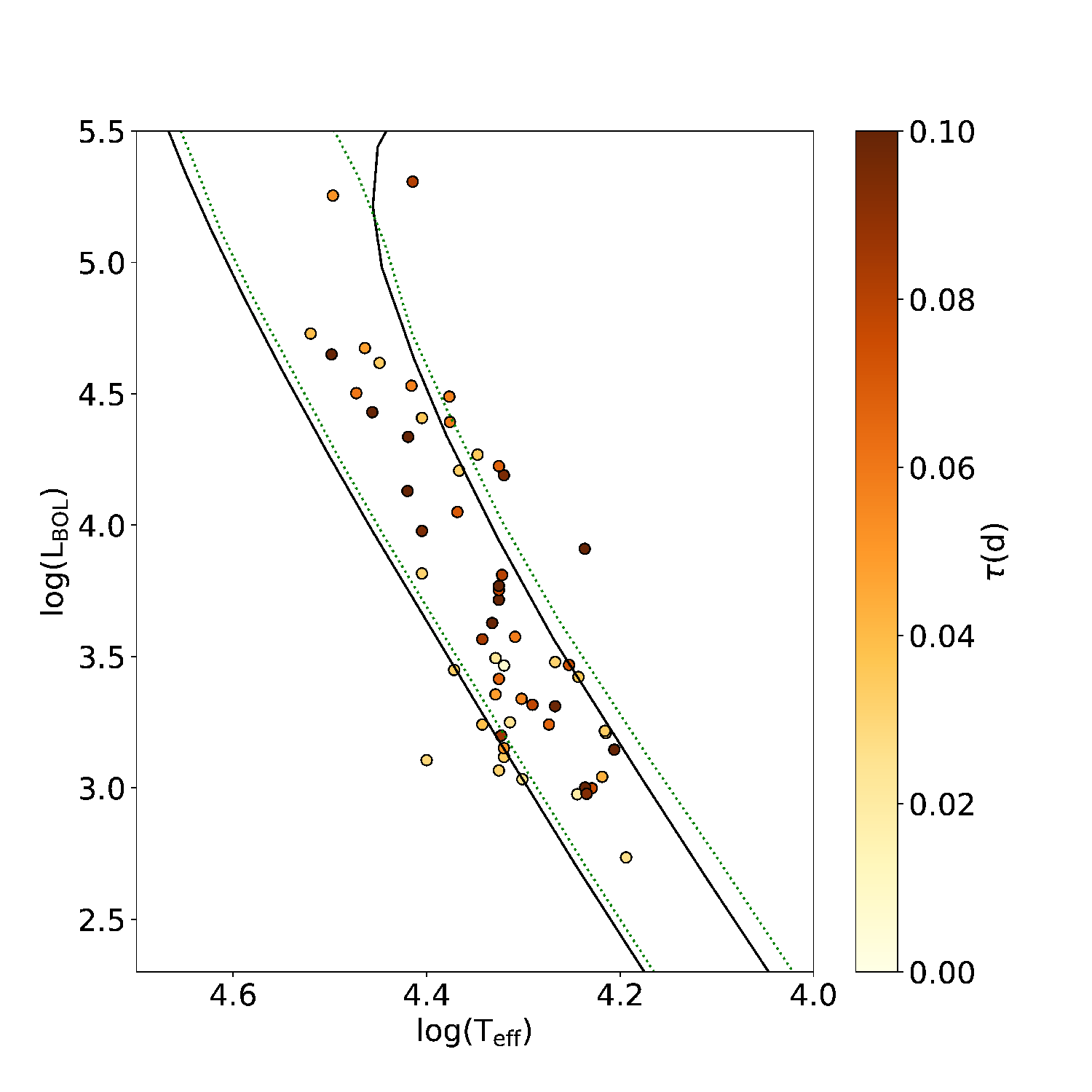}
  \caption{Hertzsprung-Russell diagram. The symbol colours indicate the value of the red noise parameters $A_0$ and $\tau$. The solid black and dotted green lines are the zero-age and terminal-age main sequences (ZAMS and TAMS) from Geneva stellar evolution models for solar abundance without and with rotation, respectively \citep{eks12}. 
  }
  \label{hrd}
\end{figure}

To search for correlations between noise parameters, we calculated the Pearson coefficients (see also Fig. \ref{cmag}) but none indicated a strong correlation ($>90$\%) -- the largest value, 44\%, was found for the $A_0-C$ relationship. Strong correlations with projected rotational velocity, effective temperature, or bolometric luminosities were also not obvious -- the largest value is only 44\%, and it was found for $\log(A_0)-\log(L_{\rm BOL})$ pairs. Finally, stars were plotted in the Hertzsprung-Russell diagram, with symbols coloured as a function of red noise parameters. While \citet{bow20} found that the red noise amplitude $A_0$ and the lifetime $\tau=1/(2\pi\nu_{\rm char})$ decreased towards the zero-age main sequence (ZAMS; i.e. higher temperatures and lower luminosities), we find no such correlation here (Fig. \ref{hrd}). However, the stellar parameters of our targets may be less precisely constrained as they often are less studied objects (see Section 2). Our conclusion could therefore benefit from a confirmation after an in-depth spectroscopic investigation of all targets, taking into account the impact of their fast rotation, become available.

\begin{figure}
  \centering
  \includegraphics[width=9cm]{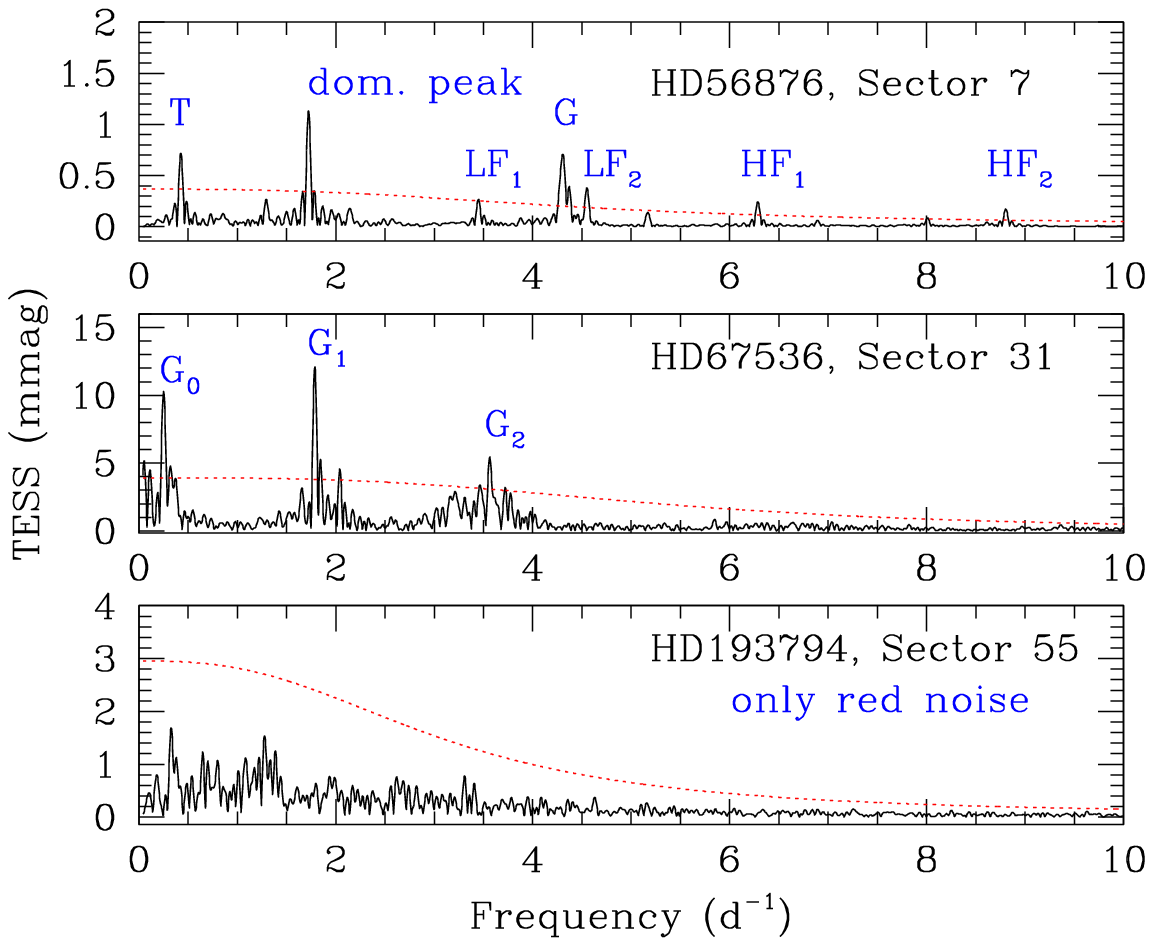}
  \caption{Three examples of frequency spectra. HD\,56876 displays red noise, long-term variations (peak labelled `T' at $f<0.5$\,d$^{-1}$), isolated signals at low and high frequencies (labelled `dom. peak', $LF_i$, $HF_i$), and a small FG (labelled `G'). HD67536 displays FGs (labelled $G_i$) while HD193794 only displays red noise. The dashed curves correspond to the significance level, set to five times the noise level. Figures showing the full set of frequency spectra are available on Zenodo (see footnote 11). }
  \label{fourier}
\end{figure}

\section{Other variations}
Figure \ref{fourier} displays examples of frequency spectra, showing the main features that are discussed in this paper: red noise, long-term variations, frequency groups (FGs), and isolated peaks. Signals are considered significant if their amplitude is five times larger than the best-fit red+white noise model computed at that frequency (i.e. $S/N=5$, as recommended by \citealt{bar15}). For \te, if data from several sectors are available, it may happen that a signal appears at the same frequency in all sectors but is formally significant in only a subset of them: such signals were also added to the list of detected features. It may be noted that only 6 targets (HD75869, HD75869, HD125238, HD166197, HD193794, and LS III +57 89, or 10\% of the sample) do not show any significant variability beyond white+red noise.

\begin{table*}
  \scriptsize
  \caption{Properties of strong FGs detected in our sample.
 \label{fg}}
  \begin{tabular}{llll}
    \hline
Name            & f($G_1$) in d$^{-1}$  & f($G_2$) in d$^{-1}$  & f($G_{>2}$) in d$^{-1}$ \\
                & (amplitude)           & (amplitude)           & (amplitude)             \\
\hline
NGC 869 133	& 2.65 (2.7--3.2\,mmag) & 5.34 (1.1--1.5\,mmag) \\ 
BD+62 657	& 0.96 (1.1--1.5\,mmag) \\
HD34748	        & 2.24 (0.6--0.7\,mmag) & 4.19 (0.3--0.5\,mmag) & 8.50 (0.2--0.3\,mmag) \\
HD249845	& 1.76 (6.8--11.8\,mmag)& 3.45 (4.2--8.8\,mmag) \\
HD254042	& 1.51 (2.4--3.3\,mmag) & very low amplitude \\
HD259865	& 1.39 (2.0--2.4\,mmag) & 2.66 (1\,mmag)        & $G_3$ very faint \\
HD46994	        & 1.70 (4\,mmag)        & 3.90 (9.0--13.5\,mmag)& 8.10 (1.5\,mmag) \\
HD47360	        & 1.68 (8.1--10.5\,mmag) \\
HD52463	        & 1.71 (0.4--1.0\,mmag)  & 4.04 (0.7--1.4\,mmag) & 5.65 (0.3--0.4\,mmag)$^a$, 7.36 (0.2--0.3\,mmag)$^a$, 11.4 (0.2\,mmag)$^a$ \\ 
HD56876	        & 4.31 (0.7\,mmag) \\
HD67536	        & 1.79 (8.9--13.5\,mmag)& 3.60 (3.6--7.4\,mmag) \\
HD68217	        & 1.93 (1.3--2.7\,mmag) \\
HD68962	        & 2.34 (1.5--2.0\,mmag) & 4.77 (5.5--7.0\,mmag) & 9.40 (0.5--1\,mmag), 14.2 (0.5--0.9\,mmag), 23.5 (0.06--0.18\,mmag), 28.1 (0.07--0.08\,mmag), 33.0 (0.05--0.06\,mmag) \\
HD81347	        & 0.76 (1.3--1.6\,mmag) & 1.55 (0.8--1.0\,mmag) & 2.31 (0.1--0.3\,mmag), 4.55 (0.07\,mmag) \\
HD93501	        & 0.96 (4.5--9.5\,mmag) & 1.86 (3.2--9.7\,mmag) \\ 
HD108257	& 2.92 (0.9--1.0\,mmag) & 4.80 (2.7--3.0\,mmag) & 6.71 (0.1--0.2\,mmag), 9.60 (0.1\,mmag), 14.3 (0.2\,mmag) \\
HD142378	& 1.89 (1.6\,mmag) \\
CPD-50 9216	& 1.37 (4.8--7.4\,mmag) & 2.69 (3.1\,mmag) \\
HD168905	& 1.88 (1.--1.4\,mmag)  & 3.89 (0.3--0.4\,mmag) \\
HD216092	& 3.10 (2.1--3.2\,mmag) & 6.05 (1.2--2.4\,mmag) \\
NGC 7654 485	& 1.96 (0.6--1.0\,mmag) \\
HD223145	& 1.39 (1.1--1.4\,mmag) & 2.77 (0.2--0.3\,mmag)$^b$ & 3.94 (0.1\,mmag)$^b$ \\
    \hline
  \end{tabular}
\tablefoot{Label $^a$ indicates information valid for Sectors 6 and 7 only, $^b$ for Sectors 1 and 2 only. Due to the variable nature of these FGs, frequencies and amplitudes should be considered approximate.}
\end{table*}

\subsection{Long-term variations ($f<0.5$\,d$^{-1}$)}
As in \citet{lab22}, we defined long-term variations as signals detected with $>2$\,d timescales. However, the duration of the {\it Kepler} observations and the duration of one \te\ sector are limited, hence very slow variations (with timescales $>30$\,d) cannot be easily identified. We thus focused on long-term variations occurring on timescales of a few days to a few weeks.

In Be stars, `flickers' are often detected (in about one-third of early-type stars observed in the first year of the \te\ mission, see \citealt{lab22}). They are characterised by their shape, with a change in brightness (increase or decrease) followed by a return to the baseline luminosity; they are not oscillations around an average luminosity. In our sample, the only variations that could maybe be related to such variability are found in HD68217 and CPD-50 9216 (so an incidence rate of maximum 2/58). Flickers are thus extremely rare amongst fast-rotating, non-Be stars. This might be understood as such variability is thought to be related to disk-building events but our stars do not display disks while Be stars do.

Two other types of long-term variability are low-frequency FGs and isolated low-frequency peaks. We defined isolated peaks as peaks in the frequency spectrum that are clearly dominating their neighbourhood, that is, at least twice larger than the next highest peak in the vicinity. Eight targets showed such isolated low-frequency peaks, with one more star showing such signals in one sector. These are:
\begin{itemize}
\item HD47360: $f$=0.140\,d$^{-1}$ (amplitude of 43--51\,mmag)
\item HD56876: $f$=0.424\,d$^{-1}$ (amplitude of 0.7\,mmag)
\item HD87152: $f$=0.264\,d$^{-1}$ (amplitude of 4.4--5.8\,mmag)
\item HD97499: $f$=0.376\,d$^{-1}$ (amplitude of 1.5--1.8\,mmag)
\item HD150745: $f$=0.416\,d$^{-1}$ (amplitude of 1.6--3.4\,mmag)
\item HD168905: $f$=0.352\,d$^{-1}$ (amplitude of 0.6\,mmag) and 0.486\,d$^{-1}$ (amplitude of 0.7--0.8\,mmag)
\item HD180968: $f$=0.174\,d$^{-1}$ (amplitude of 4.4--7.3\,mmag)
\item Cl* Berkeley 86 HG 261: $f$=0.174\,d$^{-1}$ (amplitude of 2.0--2.7\,mmag)
\item HD46883 in Sector 33: $f$=0.432\,d$^{-1}$ (amplitude of 5.2\,mmag), with its close harmonics 0.876\,d$^{-1}$ (amplitude of 5.7\,mmag) 
\end{itemize}
A range in amplitudes is provided when the amplitude is seen to change from sector to sector.

In contrast, a FG is a collection of many closely spaced peaks with several showing similar amplitudes. In Be stars, such groups most often appear at frequencies of 0.5--6.\,d$^{-1}$, but they may be complemented by a similar group at even lower frequencies (called `$G_0$' in \citealt{lab22}). This is also the case here: $G_0$ groups are not found alone. Four cases were spotted amongst our targets. It must be noted that, as low frequencies are always affected by red noise, it may become difficult to distinguish a low-amplitude FG from strong red noise. In five additional cases, the $G_0$ groups should thus be considered as tentative (hence the question marks in the list below). In the following list, the frequencies and amplitudes are those of the highest peak of the group, usually close to group centre. As such FGs are quite variable from one sector to the next, they are much less well defined than strong isolated peaks, of course, so that values are approximate. Cases are:
\begin{itemize}
\item HD67536: $G_0\sim0.25$\,d$^{-1}$ (max. amplitude of 4.1--15.0\,mmag)
\item HD68962: $G_0\sim0.1-0.2$\,d$^{-1}$ (max. amplitude of 1\,mmag)
\item HD93501: $G_0\sim0.19$\,d$^{-1}$ (max. amplitude of 6.7--12.0\,mmag)
\item CPD-50\,9216: $G_0\sim0.17$\,d$^{-1}$ (max. amplitude of 6.7--10.8\,mmag)
\item HD68217: $G_0\sim 0.16$\,d$^{-1}$ (max. amplitude of 6.3--12.4\,mmag)?
\item HD87015: $G_0\sim	0.39$ or 0.57\,d$^{-1}$ (max. amplitude of 1.0--1.4\,mmag)?
\item HD108257: $G_0\sim0.61$\,d$^{-1}$ (max. amplitude of 0.4--0.5\,mmag)?
\item NGC\,7654\,485: several peaks below 0.5\,d$^{-1}$, an ill-defined $G_0$?
\item HD223145: $G_0\sim0.34$\,d$^{-1}$ (max. amplitude of 1.6--2.3\,mmag)?
\end{itemize}

In total, there are thus 12 cases with clear low-frequency variability, plus 6 good candidates. However, this low-frequency variability clearly dominates the frequency spectrum (i.e. has a larger amplitude than any other signal) in only seven cases: HD46883 (Sect. 33), HD47360, HD87152, CPD-50\,9216, HD150745, Cl* Berkeley 86 HG 261, and HD223145. This represents a 12\% incidence rate, while \citet{lab22} counts 51\% of `longer-term trends dominating' cases in their \te\ sample of early Be stars and \citet{lab17} noted that about half of their early Be sample observed with {\it KELT} displays long-term variability or periodicities of 2--200\,d$^{-1}$. Definitely, be it for flickers or other variabilities, the fast-rotating stars without (known) disks appear much less variable on long timescales. Disk built-up and disappearance most probably explain this difference.

As a final note on long-term changes, we can also compare the frequency spectra of the same targets taken at different times (i.e. in different sectors). Compared to what is seen in Be stars \citep{lab22}, the spectra of our targets appear overall more stable, or rarely with radical changes in appearance. One exception is HD46883 which appears much more variable (stronger signals at low frequencies) in Sector 33 compared to Sector 6. More precisely, the star displays a larger variability for a few cycles around Julian date 2\,459\,208. 

\subsection{Frequency groups with $f>0.5$\,d$^{-1}$}
As noted in previous section, typical FGs occur in the 0.5--6.\,d$^{-1}$ range. In previous observations of early-type stars, FGs were found with various amplitudes. This is also the case here. Indeed, several targets of our sample (e.g. HD5882, HD35532, HD252214, HD53755, HD68324, HD78548, HD87015, HD97499, HD97913, CPD-48 8710, HD180968, HD198781, and HD207308) display low-level FGs (i.e. FGs much fainter than other signals and often below significance). In contrast, 22 of our targets (i.e. about one-third of the sample; see Table \ref{fg}) display FGs that can be qualified as `strong', that is, they dominate their frequency spectrum in the low-frequency range. This is in line with statistics from early Be stars \citep{bal20, naz20}, although FGs in such stars are often even more impressive (e.g. broader, even more complex groups).

Of our 22 targets with FGs, only six (or 27\% of the FG cases) show a single, strong group in the 0.5--6.\,d$^{-1}$ range. The majority of cases actually display at least two groups, usually noted $G_1$ and $G_2$. Frequencies of $G_1$ groups range between 0.8 and 3.1\,d$^{-1}$ while those of $G_2$ groups are about twice larger (80\% of cases have $f(G_2)/f(G_1)$ between 1.9 and 2.1). Again, this is fully in line with results obtained on Be stars \citep{lab22}. A difference with Be stars can be found when comparing amplitudes of the groups. In our sample, the $G_1$ groups display larger amplitudes than the $G_2$ groups in two-thirds of stars, while the remaining third mostly have a larger amplitude for $G_2$ groups ($G_1$ and $G_2$ groups with similar amplitudes are very rare). Be stars rather seem to split evenly between the three possibilities ($A(G_1)>A(G_2)$, $A(G_1)\sim A(G_2)$, $A(G_1)<A(G_2)$), with amplitude changes during flickers. It may also be interesting to note that additional groups are detected in 8 targets (33\% of FG cases), with one case (HD68962) displaying groups up to extremely high frequencies. Finally, r-modes have been proposed to exist in fast-rotating stars \citep{sai18}. They can be found in a typical `hump and spike' morphology in which a FG associated with the retrogrades r-modes (the `hump') is complemented by a strong peak at its high-frequency limit (the `spike', possibly corresponding to the rotational frequency). While a few cases are known in Be stars  \citep{naz20,lab22}, this morphology appear overall quite rare and this is also the case of our sample, for which none of the detected FGs display it.

In slowly pulsating B-type (SPB) stars (usually later than B3) and in $\gamma$\,Dor stars (mid-A to early F), FGs are commonly interpreted as due to multi-mode pulsations \citep{van15,van16,pap17,ped21}. But in early B-stars, which are outside the standard SPB instability strip (as calculated for non-rotating stars), the origin of FGs is more debated. They may be considered as a consequence of fast rotation \citep{sai18b,lee20}, of changes in the circumstellar environment (i.e. decretion disks of Be stars, \citealt{baa16}), or of rotation of an inhomogeneous surface \citep{bal20,bal21}. Since our sample consists of stars that are fast-rotating but not known as Be stars (i.e. not known to have circumstellar disks), the second scenario appear less plausible. However, differences do exist with Be stars, notably regarding the amplitudes of the groups; hence, the circumstellar environment may possibly have a second-order impact, not on the presence of the groups themselves but on their morphologies.

\subsection{Isolated signals with $f>0.5$\,d$^{-1}$}
Isolated peaks are typical of stellar pulsations but they could also be related to rotation\footnote{The fundamental frequency is however limited by the critical rotation rate, about 3\,d$^{-1}$ for the targets.} if the stellar surface is asymmetric (e.g. spots). Such peaks are also often detected in the frequency spectra of our targets (42 cases, or 72\% of the sample; see the list in Table \ref{domf}). By `isolated', we mean that these peaks clearly dominate their neighbourhood, that is to say, there are only much fainter peaks around, or no other peaks at all, beyond aliases of the main signal. It is usual to split such signals into low-frequencies (0.5--6.\,d$^{-1}$) and high frequencies (6.--15.\,d$^{-1}$). This splitting is related to the nature of the pulsations, with low-frequencies most probably g-modes and high frequencies p-modes.

Nearly two-thirds of our sample (37 stars, or 64\%) display isolated signals at low frequencies (0.5--6.\,d$^{-1}$). Such peaks may have low amplitudes but, in 24 cases (41\% of the sample), the strongest signal recorded in the frequency spectrum is a peak, or a pair of peaks, in this range (hence our choice of a `dominant' qualifier; see Col. 2 of Table \ref{domf}). Four additional stars (HD46883, HD87152, HD150745, and Cl* Berk. 86 HG 261) display a stronger signal at even lower frequencies (below 0.5\,d$^{-1}$) but isolated peaks in 0.5--6.\,d$^{-1}$ dominate the spectra above 0.5\,d$^{-1}$, increasing the incidence of such dominant low-frequency signals to 48\%. This is much larger than the fraction of 22.5\% of early-type Be displaying isolated, low-frequency signals found by \citet{lab22} and is still above the larger, 20--40\%, incidence found in \citet{see18}, \citet{bal20}, or \citet{naz20}.

Signals at higher frequencies (6.--15.\,d$^{-1}$) are rarer: they are detected in 20 cases (34\% of the sample). Such peaks are generally fainter than those at lower frequencies, with only 3 stars (HD68324, HD143118, and HD192968) having the largest-amplitude signals at high frequencies and 2 other targets (HD37303 and HD207308) displaying low and high-frequency peaks with similar amplitudes. This recalls the fraction of $\sim10$\% of early Be stars with strong high-frequency signals reported by \citet{naz20} and \citet{bal20}. Signals with even higher frequencies (15.--40.\,d$^{-1}$) are found only in 11 cases (19\% of the sample), again larger than the incidence of 2\% found by \citet{lab22}. Usually, such very high-frequency signals are quite faint: there is only one case (HD37303) with peaks of similar strengths at low, high, and very high frequencies.

The detected signals may not be independent. About 30\% of the stars with isolated signals display harmonics (slightly fewer than one-half if taking close harmonics into account). For example, HD37397 displays several multiples (2, 4, 6, 7, and 9 $\times$) of the dominant signal at 1.744\,d$^{-1}$, as for HD35777 (2, 4, 7, 9, 12, and 15 $\times$ the dominant signal at 1.716\,d$^{-1}$), or HD37303 (2, 7, 9, and 16 $\times$ the dominant signal at 2.140\,d$^{-1}$). This is quite normal as, when using Fourier techniques, harmonics will naturally appear whenever the variability is not perfectly sinusoidal. It usually is more typical of rotational (e.g. \citealt{dav19}) or binarity signals than of pulsations \citep{lab20}, though.  Combination of frequencies can also be found in a few cases: in HD37674, the difference between 2.764 and 3.304\,d$^{-1}$ is similar to that between 6.076 and 6.612\,d$^{-1}$ or 12.144 and 12.664\,d$^{-1}$; in HD136298, the difference between 2.538 and 2.700\,d$^{-1}$ is also found between 4.710 and 4.870\,d$^{-1}$ while that between 8.580 and 9.770\,d$^{-1}$ resembles that between 10.576 and 11.746\,d$^{-1}$.

\begin{figure}
  \centering
  \includegraphics[width=9cm]{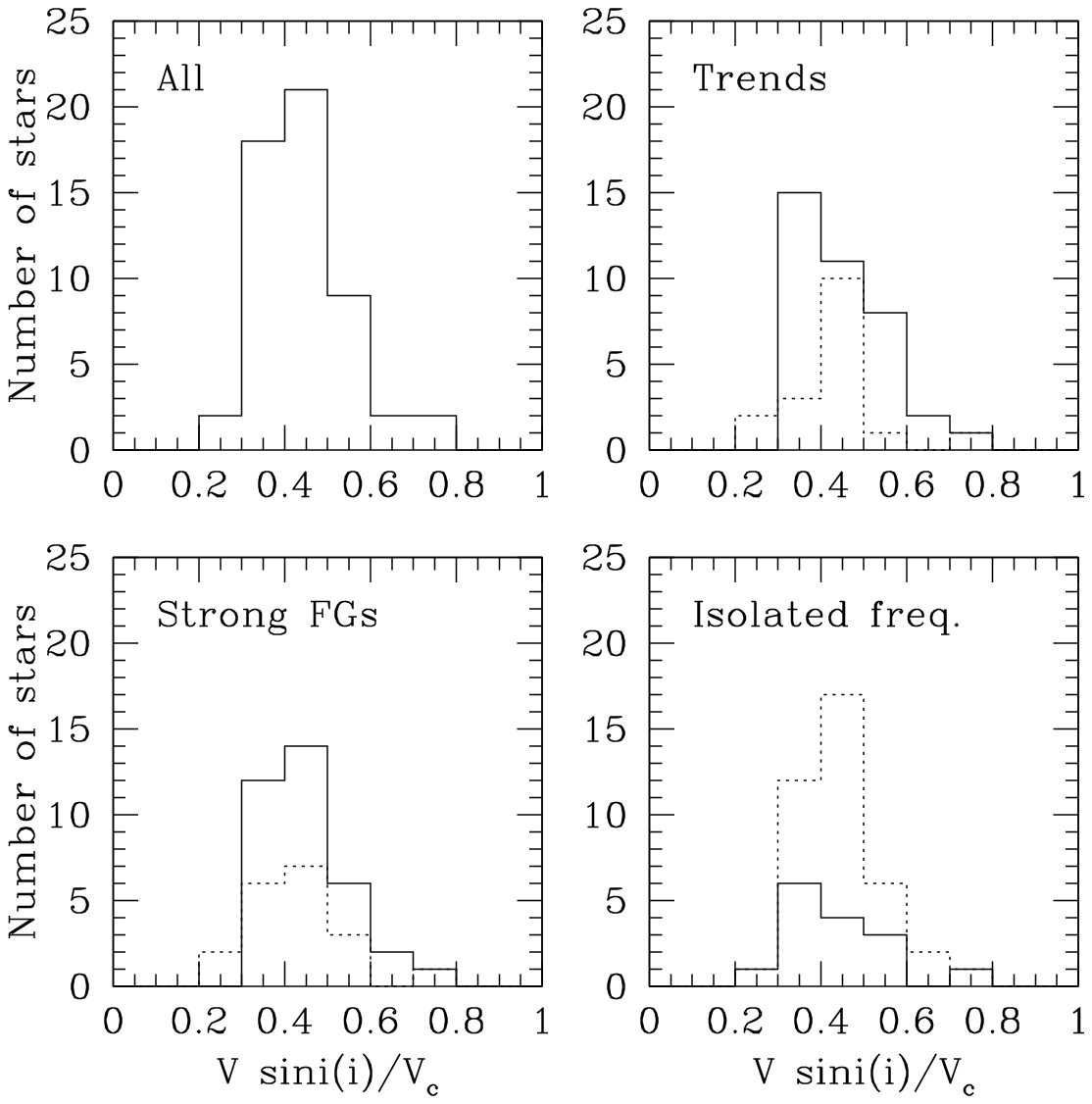}
  \caption{Histograms of ratios $v\sin(i)/V_c$ for all targets (top-left panel) or targets separated by the type of signal recorded (trends on top right panel, strong FGs on bottom left panel, and isolated frequencies on bottom right panel). Cases with and without features are shown by dotted and solid lines, respectively.  
  }
  \label{histo2}
\end{figure}

\begin{figure}
  \centering
  \includegraphics[width=9cm]{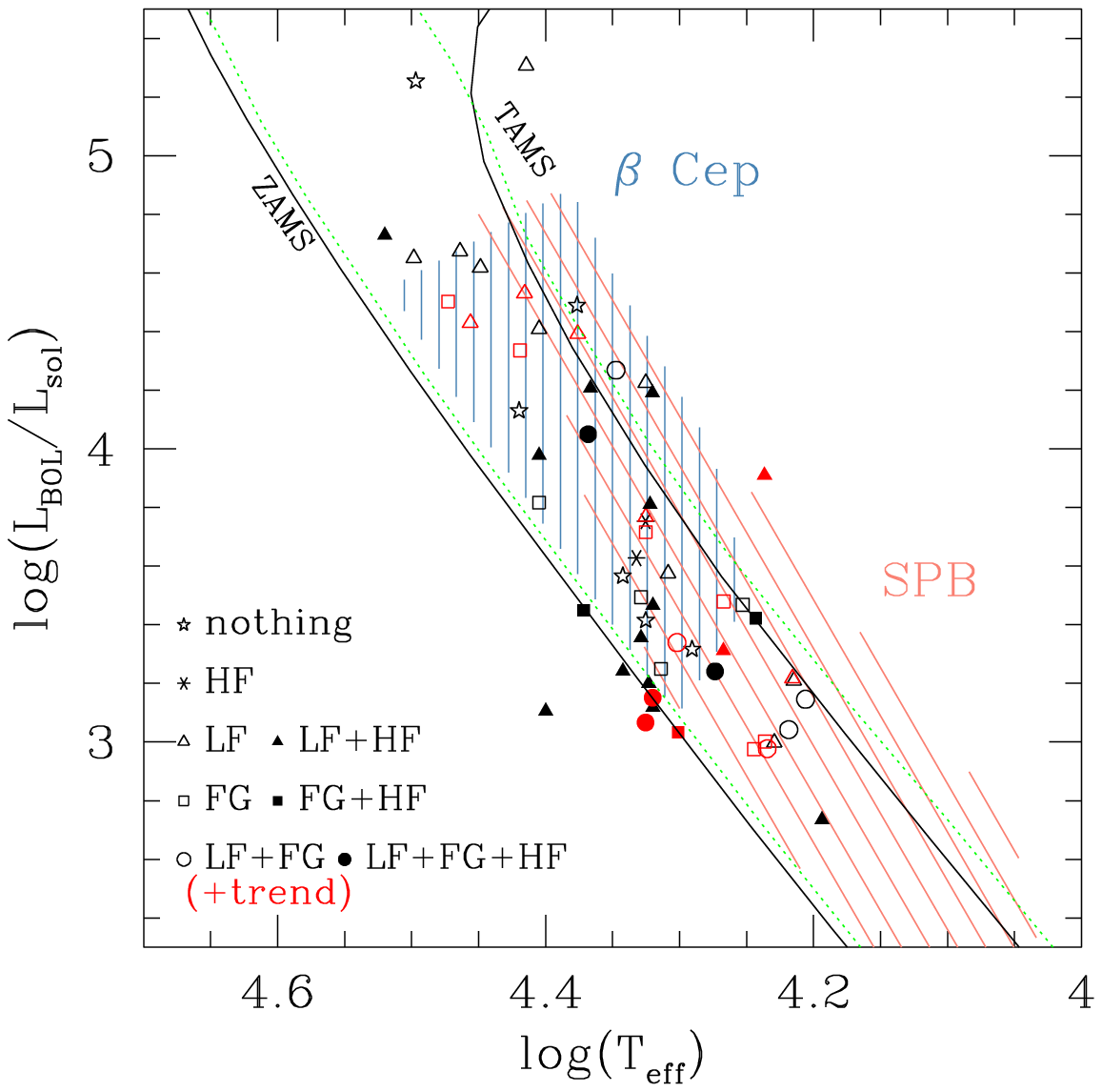}
  \caption{Hertzsprung-Russell diagram. The symbols depict the type of variability and have the same meanings as in Fig. \ref{cmag}. The solid black and dotted green lines are the ZAMS and TAMS from Geneva stellar evolution models for solar abundance without and with rotation, respectively (\citealt{eks12};  see our Fig. \ref{hrd}). The hatched areas mark the instability zones of $\beta$\,Cep and SPB stars from \citet{mig07}. 
  }
  \label{hrd2}
\end{figure}

Figure \ref{histo2} shows histograms of observed criticality ratios $v\sin(i)/V_c$ for all targets as well as targets displaying (or not) some feature. As can be seen, there is no significant difference between the samples with or without the chosen features (trends, strong FGs, isolated signals). No physical correlation between one feature and rotation can thus be derived.  Figure \ref{hrd2} shows the targets, with symbols coding their variability type, in a Hertzsprung-Russell diagram with instability zones indicated. Targets appear between the ZAMS and the TAMS, as could be expected. Targets with high-frequency signals (asterisk or filled symbols) do not appear clearly separated from targets with only low-frequency signals (open symbols). High-frequency pulsators are even found far out of the $\beta$\,Cep area. Of course, uncertainties on temperatures and bolometric luminosities can be expected as fast rotation and unknown viewing angle could bias the estimates of these quantities (see also Section 2). In addition, detailed asteroseismic modelling would be required to better understand the behaviour of these objects.

\section{Conclusions}
Fast-rotating OB stars come in two flavours, those with disks (also known as Oe/Be stars) and those without disks. The variability of the former group has been repeatedly examined over the years, while the latter group benefited from less attention. Using data from \te\ and {\it Kepler}, we report here a detailed photometric study of 58 isolated, early-type ($>$B3), fast-rotating ($v \sin(i)>200$\,\kms)  massive stars not known to  display disks. This work aims at identifying the similarities and differences in the variability patterns of rapidly rotating early B-type stars with and without the Be phenomenon. The difference in behaviour may provide clues as to which variability is typical of circumstellar matter and which is not, providing guidance for future modelling.

The presence of white+red noise is ubiquitous in our sample, as in OB stars in general. Telescope sensitivity limits the detection of faint white noise, and red noise always appear stronger than white noise in our targets. Correlations between different red noise parameters and between these parameters and stellar properties were searched for, but none could be found.

Signals significantly above the noise level are detected in 90\% of the sample but such variability occurs on very different timescales. First, 20--30\% of stars in our sample are affected by long-term changes, which even dominate the frequency spectra for 12\% of the total sample. There is clearly a a lower incidence of long-term variability, caused by isolated peaks or very low-frequency frequency groups ($G_0$), in our sample than in Be stars, underlining the role of circumstellar changes on long-term variability.

{\it Strong} FGs are detected in 38\% of our sample, often with two or more groups forming a near-harmonic series (e.g. $f(G_2)\sim 2f(G_1)$), as in Be stars. In other words, the occurrence of FGs may somehow be linked to fast rotation but it is clearly {\it not} sufficient to classify a star as a Be type since FGs can also be found in fast-rotating non-Be stars. Moreover, this also implies that FGs do not appear to be a crucial factor for the Be phenomenon (i.e. for the creation of a disk). However, in our sample, the first FG most often displays the largest amplitude, unlike in Be stars. This difference in the relative strengths of FGs in the Be versus non-Be stars suggests that, while not being fully dependent on each other, some interplay exists between the circumstellar environment and these groups.

Isolated signals are frequent in our sample, especially in the 0.5--6.\,d$^{-1}$ range: two-thirds of our stars display such low-frequency signals and these signals even dominate the frequency spectrum for 41\% of our sample. Both of these rates are higher than what is found for Be stars. Higher-frequency signals also are more frequent in our sample than in Be stars, although they are rarely dominating (this concerns only 9\% of our cases). It should be noted that the presence of high-frequency signals is not limited to stars in the $\beta$\,Cep locus. 

While variability patterns are now identified, much remains to be done. First, our conclusions would be strengthened by an in-depth spectroscopic analysis of our sample stars. High-quality spectra, coupled with atmosphere modelling of distorted stars, would allow us to determine the stellar properties more precisely. A sparse monitoring of our targets could also be useful, to assess whether some of them display emission lines at some point (the Be phenomenon may be intermittent). Finally, high-cadence spectroscopic monitoring would further allow us to identify the spectroscopic variability features associated to the photometric changes. The next step would then be to perform a detailed asteroseismic modelling of stars with and without disks using the derived constraints.

Overall, one still needs to identify what makes a Be star so special. While one could expect a strong impact on its behaviour due to the presence of circumstellar material, this work demonstrates that Be and non-Be stars actually share many similarities in their photometric properties. This may not be entirely surprising as, for some Be stars, the overall properties of the frequency spectra do not seem to depend on the disk size (see Appendix D). Photometric data thus seem more sensitive to the occurrence of ejection events than the simple presence or absence of a circumstellar disk. This is particularly intriguing as the presence of Be disks has been proposed to be linked to specific pulsational behaviour (e.g. \citealt{hua09}). One additional ingredient should probably be considered, such as how strong or common mode coupling is. Fast rotation is definitely not sufficient to explain the occurrence of the Be phenomenon, and neither is the presence of photometric FGs.

\begin{acknowledgements}
The authors acknowledge the referee, J. Zorec, for his interesting comments and S; Ekstr\"om for help with her evolutionary models. Y.N. acknowledges support from the Fonds National de la Recherche Scientifique (Belgium), the European Space Agency (ESA) and the Belgian Federal Science Policy Office (BELSPO) in the framework of the PRODEX Programme (contracts linked to XMM-Newton and Gaia). N.B. acknowledges support from the Belgian federal government grant for Ukrainian postdoctoral researchers (contract UF/2022/10). This paper includes data collected by the TESS mission, which are publicly available from the Mikulski Archive for Space Telescopes (MAST). Funding for the TESS mission is provided by NASA's Science Mission directorate. This paper includes data collected by the Kepler mission and obtained from the MAST data archive at the Space Telescope Science Institute (STScI). This work makes use of observations from the Las Cumbres Observatory global telescope network. This work uses observations obtained at the Dominion Astrophysical Observatory, NRC Herzberg, Programs in Astronomy and Astrophysics, National Research Council of Canada. ADS and CDS were used for preparing this document. 
\end{acknowledgements}

\begin{appendix}

\section{Target list}

\begin{sidewaystable*}
  \scriptsize
  \caption{Target list, ordered by increasing RA.  \label{list}}
  \begin{tabular}{llcccccccccc}
    \hline
Name     & Sectors                  & TIC       & $d$(pc)         & SpT & Ref. & $v\sin(i)$ & Ref. & $T_{eff}$(K) & Ref. & $\log(L_{\rm BOL}$ & $v\sin(i)/$\\
         &                          &           &                 &     &      & (\kms)     &      &             &      & $/L_{\odot})$ &   $V_c$\\
    \hline
HD5882   & 17,18,58                 & 299328900 &  423.7$\pm$7.7  & B2.5Vn & \citet{2021AA...645L...8X}  & 358 & \citet{hua10}               & 15624 & \citet{hua10}              & 2.74 & 0.60\\
NGC 869 133            & 18,58      & 348137277 & 2555.6$\pm$279.3&        &                             & 341 & \citet{mar12}               & 17500 & \citet{mar12}              & 3.42 &     \\
BD+56 538& 18,58*                   & 348231861 & 2221.3$\pm$70.0 & B2     & \citet{2021AA...645L...8X}  & 235 & \citet{mar12}               & 20900 & \citet{mar12}              & 4.19 & 0.61\\
HD14250  & 18*,58                   & 348314700 & 2485.1$\pm$585.9& B1V    & \citet{2010ApJS..186..191C} & 236 & \citet{2006ApJ...648..591H} & 25972 & \citet{2006ApJ...648..591H}& 5.31 & 0.77\\
BD+62 657& 19,59*,73*               & 84429626  & 1015.0$\pm$18.2 & B2V    & \citet{2021AA...645L...8X}  & 212 & \citet{2006ApJ...648..591H} & 23347 & \citet{2006ApJ...648..591H}& 4.05 & 0.46\\
HD34748  & 5,32*                    & 4011607   &  358.8$\pm$11.2 & B1.5V  & \citet{2005AA...444..941P}  & 295 & \citet{coc20}               & 23529 & \citet{coc20}              & 3.45 & 0.38\\
HD35532  & 6*,32*,43*,44*,45*,71*   & 302267223 & 333.6$\pm$7.9   & B2Vn   & \citet{abt02}               & 281 & \citet{hua10}               & 16397 & \citet{hua10}              & 3.21 & 0.53\\
HD35777  & 6,32*                    & 50524860  &  344.3$\pm$5.9  & B2V    & \citet{2010AN....331..349H} & 230 & \citet{bra12}               & 22000 & \citet{2010AN....331..349H}& 3.24 & 0.33\\
HD37303  & 6*,32*                   & 332856560 &  362.9$\pm$9.8  & B1.5V  & \citet{2020AA...639A..81B}  & 280 & \citet{2020AA...639A..81B}  & 20893 & \citet{2020AA...639A..81B} & 3.47 & 0.41\\
HD37397  & 6,32                     & 11199427  &  345.8$\pm$6.6  & B2V    & \citet{bra12}               & 236 & \citet{bra12}               & 20893 & \citet{2005AJ....129..856H}& 3.12 & 0.33\\
HD37674  & 6                        & 11299576  &  404.5$\pm$6.5  & B5Vn   & \citet{1976AJ.....81..537G} & 385 & \citet{2005csss...13..571G} & 25119 & \citet{2005AJ....129..856H}& 3.10 &     \\
HD249845 &19,43*,44*,45*,71*,72*,73*& 353298304 &  796.5$\pm$23.6 & B2:V:nn& \citet{2012AstL...38..331A} & 218 & \citet{2005csss...13..571G} & 20600$^m$ &                        & 3.25 & 0.34\\
HD252214 & 6,33,43*,44*,71*,72*     & 59468295  &  920.7$\pm$29.3 & B2.5V  & \citet{2012AstL...38..331A} & 280 & \citet{2005csss...13..571G} & 20350 & \citet{2010AN....331..349H}& 3.57 & 0.58\\
HD254042 & 71,72                    & 83017766  & 1629.9$\pm$54.7 & B0.5:IV:nn &\citet{xia22}               &226 & \citet{xia22}             & 22253 & \citet{xia22}              & 4.27 & 0.40\\
HD259865 & 6,33                     & 234951915 & 1311.8$\pm$31.7 & B5V    & \citet{2014ASPC..485..223B} & 256 & \citet{2006ApJ...648..591H} & 18773 & \citet{2006ApJ...648..591H}& 3.24 &     \\
HD46994  & 6*,7*,33*                & 172233421 &  860.7$\pm$30.0 & B2/3V  & \citet{bra12}               & 271 & \citet{bra12}               & 21320 & \citet{bra12}              & 3.49 & 0.51\\
HD46883  & 6,33                     & 220134169 & 1123.3$\pm$34.4 & B0.5:V & \citet{2018AA...616L..15X}  & 227 & \citet{2005csss...13..571G} & 23764 & \citet{2007ApJ...663..320F}& 4.39 & 0.41\\
HD47360  & 6,33                     & 319854134 & 1319.9$\pm$57.4 & B0.5V  & \citet{2020MNRAS.493.2339M} & 237 & \citet{daf07}               & 26250 & \citet{daf07}              & 4.34 & 0.37\\
HD52463  & 6,7,33,34                & 63359889  &  831.1$\pm$22.3 & B3V    & \citet{bra12}               & 262 & \citet{bra12}               & 16540 & \citet{bra12}              & 3.04 & 0.52\\
HD53755  & 7,33*                    & 177129056 & 1050.4$\pm$43.3 & B0.5V  & \citet{2003AN....324..219W} & 285 & \citet{2017AA...603A..56C}  & 28100 & \citet{2017AA...603A..56C} & 4.62 & 0.49\\
HD56876  & 7*,33*,34*,61*           & 66594335  &  279.7$\pm$3.0  & B2IV   & \citet{2010AN....331..349H} & 310 & \citet{abt02}               & 21150 & \citet{2010AN....331..349H}& 3.07 & 0.42\\
ALS 864  & 7,8,34,61                & 129364127 & 4054.6$\pm$399.1& B0V    & \citet{1981AAS...45..193P}  & 249 & \citet{2017AA...603A..56C}  & 31500 & \citet{2017AA...603A..56C} & 4.65$^m$ & 0.36\\
HD67536  & 1,4*,7*,8*,9*,10*,11,    & 308542383 & 419.9$\pm$7.3   & B2.5Vn & \citet{2005ApJS..158..193S} & 325 & \citet{2005csss...13..571G} & 18500$^m$ &                        & 3.48 & 0.70\\
         & 28*,31*,34*,35*,37*,38*, \\
         & 61*,62*,63*,64*,68*,69*\\
HD68217  & 7*,8*,34*,35*,61*,62*    & 354931043 &  329.6$\pm$15.9 & B2IV   & \citet{2010AN....331..349H} & 214 & \citet{2005csss...13..571G} & 21150 & \citet{2010AN....331..349H}& 3.72 & 0.42\\
HD68324  & 7*,8*,9*,34*,35*,61*,62* & 238612881 &  342.5$\pm$8.9  & B2IV   & \citet{2005ApJS..158..193S} & 210 & \citet{2005csss...13..571G} & 21150 & \citet{2010AN....331..349H}& 3.75 & 0.42\\
HD68962  & 7*,8*,34*,35*,61*        & 182315791 &  460.9$\pm$7.6  & B2/3V  & \citet{bra12}               & 293 & \citet{bra12}               & 17550 & \citet{bra12}              & 2.98 & 0.50\\
HD75869  & 8,35,62                  & 190264184 &  589.5$\pm$12.4 & B2V    & \citet{2010AN....331..349H} & 208 & \citet{1975MmRAS..78...51B} & 22000 & \citet{2010AN....331..349H}& 3.57 & 0.36\\
HD78548  & 8,10,35*,37*,62*,63*     & 384642833 &  342.1$\pm$8.8  & B2IV   & \citet{2010AN....331..349H} & 215 & \citet{1975MmRAS..78...51B} & 21150 & \citet{2010AN....331..349H}& 3.41 & 0.36\\
HD81347  & 9,35,36,62,63            & 295691515 &  479.1$\pm$9.8  & B3IV   & \citet{2010AN....331..349H} & 250 & \citet{1975MmRAS..78...51B} & 17900 & \citet{2010AN....331..349H}& 3.47 & 0.58\\
HD87015  & 21*,45*,46*,48*,72*      & 26987416  &  308.3$\pm$7.9  & B2.5IV & \citet{2006AA...452..945T}  & 215 & \citet{2006AA...452..945T}  & 16435 & \citet{hua10}              & 3.22 & 0.45\\
HD87152  & 9,10,36*,37*,63*         & 134730105 &  361.0$\pm$8.1  & B2.5V  & \citet{1996AAS..118..481B}  & 236 & \citet{1975MmRAS..78...51B} & 18500$^m$ &                        & 3.31 & 0.46\\
HD93501  & 10,11,36,37,63,64        & 391527871 & 1880.1$\pm$73.6 & B0V    & \citet{2016AJ....152..190A} & 210 & \citet{han18}               & 29700 & \citet{2016AJ....152..190A}& 4.50 & 0.30\\
HD97499  & 10,11,37,64              & 467235690 &                 & B0.5   & \citet{2020MNRAS.493.2339M} & 201 & \citet{daf07}               & 28590 & \citet{daf07}              & 4.43$^m$ & 0.31\\
HD97913  & 10,11,37,64*             & 450359616 & 2612.5$\pm$148.7& B0.5IVn& \citet{bra12}               & 326 & \citet{bra12}               & 33110 & \citet{bra12}              & 4.73 & 0.51\\
HD108257 & 10*,11*,37*              & 260646994 &  135.6$\pm$2.7  & B3Vn   & \citet{2016AJ....152...40G} & 298 & \citet{wol07}               & 20000 & \citet{2016AJ....152...40G}& 3.03 & 0.49\\
HD125238 & 11*,38*,65*              & 242497929 &  107.9$\pm$5.2  & B2.5IV & \citet{1997AA...319..811B}  & 222 & \citet{van12}               & 19525 & \citet{2010AN....331..349H}& 3.32 & 0.41\\
HD133385 & 12,38,39,65,66           & 403073918 & 1049.1$\pm$36.4 & B2Vn   & \citet{2011MNRAS.410..190T} & 257 & \citet{1975MmRAS..78...51B} & 21150 & \citet{2010AN....331..349H}& 4.22 & 0.68\\
HD136298 & 11*,38*,65*              & 148415949 &  149.8$\pm$13.6 & B1.5IV & \citet{van12}               & 225 & \citet{van12}               & 20990 & \citet{2016AA...591A.118S} & 3.81 & 0.40\\
HD138485 & KEPLER               & ktwo200194914 &  229.4$\pm$6.6  & B2Vn   & \citet{str05}               & 203 & \citet{bra12}               & 21320 & \citet{bra12}              & 3.36 & 0.32\\
HD142378 & KEPLER               & ktwo205104403 &  264.6$\pm$27.7 & B1.5Vn & \citet{2016AJ....152...40G} & 225 & \citet{str05}               & 16069 & \citet{2007AA...466..269B} & 3.15 & 0.36\\
HD143118 & 12*,65*                  &  59095516 &  131.6$\pm$7.2  & B2.5V  & \citet{van12}               & 240 & \citet{van12}               & 21490 & \citet{2016AA...591A.118S} & 3.63 & 0.49\\
CPD-50 9216 & 12,39,66              & 315368472 & 1849.3$\pm$54.5 & B0.5IV & \citet{2001KFNT...17e.409K} & 230 & \citet{daf07}               & 20040 & \citet{daf07}              & 3.34 & 0.27\\
CPD-48 8710 & 12,39,66              & 39645989  & 1135.0$\pm$28.2 & B0.5V  & \citet{2001KFNT...17e.409K} & 296 & \citet{2006ApJ...648..591H} & 21036 & \citet{2006ApJ...648..591H}& 3.20 & 0.30\\
HD150745 & 12*,39*,66*              & 420629395 &  403.5$\pm$11.8 & B2IV   & \citet{2010AN....331..349H} & 244 & \citet{2005csss...13..571G} & 21150 & \citet{2010AN....331..349H}& 3.77 & 0.49\\
HD164900 & 40*,53*                  & 462709664 &  281.8$\pm$4.1  & B3Vn   & \citet{2005AA...444..941P}  & 260 & \citet{coc20}               & 16954 & \citet{coc20}              & 3.00 & 0.49\\
HD166197 & 13,66*                   & 57997895  &  982.3$\pm$72.2 & B1V    & \citet{1996AAS..118..481B}  & 229 & \citet{2009AN....330..317H} & 23800 & \citet{2009AN....330..317H}& 4.49 & 0.51\\
HD168905 & 13*,66*                  & 89975465  &  178.0$\pm$4.1  & B2.5V  & \citet{2010AN....331..349H} & 248 & \citet{coc20}               & 20901 & \citet{coc20}              & 3.15 & 0.39\\
HD180968 & 14*,40*,54*              & 354462452 &  544.9$\pm$23.6 & B0.5IV & \citet{abt02}               & 270 & \citet{abt02}               & 26043 & \citet{2011AA...525A..71W} & 4.53 & 0.48\\
HD192968 & 14,15,41*,55*            & 11345334  &  982.6$\pm$32.4 & B1Vne  & \citet{2010AN....331..349H} & 287 & \citet{2005csss...13..571G} & 25400 & \citet{2010AN....331..349H}& 3.98 & 0.44\\
Cl* Berkeley \\
86 HG 261& 14,15,41,55              & 274068655 & 1695.6$\pm$29.8 & B1.5V  & \citet{hua06}               & 251 & \citet{2006ApJ...648..591H} & 17241 & \citet{2006ApJ...648..591H}& 3.91$^m$ & 0.58\\
HD193794 & 14,15,41*,55*            & 13116289  & 2086.0$\pm$226.3& B0V    & \citet{2005NewA...10..325Z} & 231 & \citet{2005csss...13..571G} & 31400$^m$ &                        & 5.25 & 0.48\\
HD198781 &15*,17*,18*,24*,56*,57*,58*&305255878 &  922.7$\pm$20.3 & B0.5V  & \citet{1996AAS..118..481B}  & 222 & \citet{2017AA...603A..56C}  & 29100 & \citet{2017AA...603A..56C} & 4.67 & 0.38\\
HD201819 & 15*,55*,56*              & 166354242 &  978.2$\pm$44.7 &B0.5IVn & \citet{2021AA...645L...8X}  & 215 & \citet{abt02}               & 25400 & \citet{2010AN....331..349H}& 4.41 & 0.37\\
HD207308 & 16,17,24,56,57,58        & 408100370 &  907.7$\pm$14.3 & B0.7III-IV(n) & \citet{2012AstL...38..331A} & 207 & \citet{daf07}        & 23250 & \citet{daf07}              & 4.21 & 0.37\\
LS III +57 89          & 16,17      & 343878178 & 2854.2$\pm$188.7& B1V    & \citet{wol07}               & 204 & \citet{hua10}               & 26291 & \citet{hua10}              & 4.13$^m$ & 0.33\\
HD216092 & 16,17,56,57              & 66964209  &  874.6$\pm$78.4 & B1.5Vn & \citet{2012AstL...38..331A} & 261 & \citet{2005csss...13..571G} & 25400 & \citet{2010AN....331..349H}& 3.82 & 0.39\\
NGC 7654 485 & 17,18,24,57,58       & 269518687 & 1743.5$\pm$27.2 &        &                             & 232 & \citet{hua10}               & 17225 & \citet{hua10}              & 3.00 &     \\
HD223145 & 1*,2*,28*,29*,68*,69*    & 206362352 &  180.9$\pm$2.7  & B2.5V  & \citet{2019MNRAS.485.3457B} & 240 & \citet{2019MNRAS.485.3457B} & 17163 & \citet{2019MNRAS.485.3457B}& 2.98 & 0.42\\
    \hline
  \end{tabular}
\tablefoot{A star after the sector's number indicates a high-cadence \te\ observation, $^m$ indicates a value from Mamajek's stellar calibration.  }
\end{sidewaystable*}

\section{Red noise parameters}

\begin{table*}
  \scriptsize
  \caption{Red noise parameters for all light curves. \label{rednt}}
  \begin{tabular}{llcccc}
    \hline
                Name & Sector & $C$(mmag)         & $A_0$(mmag)     & $\tau$(d)         & $\gamma$        \\
    \hline
                 HD5882 & 17  & 0.0172$\pm$0.0022 & 0.087$\pm$0.006 & 0.0369$\pm$0.0025 & 4.11$\pm$0.95 \\
                 HD5882 & 18  & 0.0143$\pm$0.0024 & 0.087$\pm$0.007 & 0.0343$\pm$0.0024 & 4.37$\pm$1.07 \\
                 HD5882 & 58  & 0.0055$\pm$0.0002 & 0.091$\pm$0.002 & 0.0262$\pm$0.0007 & 2.64$\pm$0.11 \\
            NGC 869 133 & 18  & 0.0302$\pm$0.0312 & 0.827$\pm$0.328 & 0.4285$\pm$0.4350 & 0.69$\pm$0.21 \\
            NGC 869 133 & 58  & 0.0360$\pm$0.0005 & 0.199$\pm$0.006 & 0.0366$\pm$0.0014 & 2.71$\pm$0.17 \\
              BD+56 538 & 18  & 0.0405$\pm$0.0032 & 0.384$\pm$0.026 & 0.1567$\pm$0.0157 & 1.94$\pm$0.23 \\
              BD+56 538 & 58* & 0.0133$\pm$0.0001 & 0.296$\pm$0.004 & 0.0949$\pm$0.0017 & 2.53$\pm$0.07 \\
                HD14250 & 18* & 0.0289$\pm$0.0005 & 1.142$\pm$0.015 & 0.0809$\pm$0.0017 & 1.76$\pm$0.03 \\
                HD14250 & 58  & 0.0223$\pm$0.0007 & 0.904$\pm$0.015 & 0.0776$\pm$0.0020 & 1.83$\pm$0.04 \\
              BD+62 657 & 19  & 0.0273$\pm$0.0018 & 0.198$\pm$0.009 & 0.0799$\pm$0.0039 & 3.95$\pm$0.57 \\
              BD+62 657 & 59* & 0.0092$\pm$0.0002 & 0.158$\pm$0.004 & 0.0663$\pm$0.0030 & 1.42$\pm$0.04 \\
              BD+62 657 & 73* & 0.0095$\pm$0.0002 & 0.226$\pm$0.006 & 0.0703$\pm$0.0037 & 1.28$\pm$0.04 \\
                HD34748 &  5  &-0.0078$\pm$0.0108 & 0.109$\pm$0.020 & 0.0315$\pm$0.0052 & 1.26$\pm$0.35 \\
                HD34748 & 32* & 0.0028$\pm$0.0001 & 0.083$\pm$0.001 & 0.0337$\pm$0.0010 & 1.70$\pm$0.04 \\
                HD35532 &  6* & 0.0102$\pm$0.0003 & 0.419$\pm$0.005 & 0.0302$\pm$0.0005 & 2.65$\pm$0.07 \\
                HD35532 & 32* & 0.0038$\pm$0.0002 & 0.160$\pm$0.002 & 0.0324$\pm$0.0005 & 3.95$\pm$0.17 \\
                HD35532 & 43* & 0.0034$\pm$0.0002 & 0.162$\pm$0.002 & 0.0281$\pm$0.0005 & 3.08$\pm$0.12 \\
                HD35532 & 44* & 0.0036$\pm$0.0002 & 0.175$\pm$0.002 & 0.0284$\pm$0.0005 & 3.07$\pm$0.11 \\
                HD35532 & 45* & 0.0037$\pm$0.0002 & 0.198$\pm$0.003 & 0.0304$\pm$0.0006 & 2.71$\pm$0.10 \\
                HD35532 & 71* & 0.0042$\pm$0.0003 & 0.280$\pm$0.004 & 0.0292$\pm$0.0005 & 2.62$\pm$0.08 \\
                HD35777 &  6  & 0.0259$\pm$0.0019 & 0.087$\pm$0.005 & 0.0366$\pm$0.0010 & 18.5$\pm$8.14 \\
                HD35777 & 32* & 0.0040$\pm$0.0001 & 0.102$\pm$0.001 & 0.0384$\pm$0.0002 & 23.9$\pm$2.46 \\
                HD37303 &  6* & 0.0025$\pm$0.0001 & 0.022$\pm$0.001 & 0.0148$\pm$0.0009 & 1.40$\pm$0.07 \\
                HD37303 & 32* & 0.0028$\pm$0.0001 & 0.012$\pm$0.001 & 0.0067$\pm$0.0001 & 4.32$\pm$0.27 \\
                HD37397 &  6  & 0.0554$\pm$0.0041 & 0.244$\pm$0.011 & 0.0391$\pm$0.0006 & 41.3$\pm$22.0 \\
                HD37397 & 32  & 0.0159$\pm$0.0013 & 0.243$\pm$0.007 & 0.0345$\pm$0.0008 & 5.74$\pm$0.63 \\
                HD37674 &  6  & 0.0412$\pm$0.0057 & 0.145$\pm$0.014 & 0.0300$\pm$0.0026 & 4.55$\pm$1.50 \\
               HD249845 & 19  & 0.0691$\pm$0.0239 & 0.923$\pm$0.057 & 0.0294$\pm$0.0017 & 3.60$\pm$0.65 \\
               HD249845 & 43* & 0.0127$\pm$0.0010 & 1.210$\pm$0.015 & 0.0321$\pm$0.0005 & 2.86$\pm$0.08 \\
               HD249845 & 44* & 0.0111$\pm$0.0009 & 1.020$\pm$0.012 & 0.0317$\pm$0.0005 & 2.97$\pm$0.09 \\
               HD249845 & 45* & 0.0143$\pm$0.0009 & 1.300$\pm$0.014 & 0.0338$\pm$0.0005 & 2.85$\pm$0.07 \\
               HD249845 & 71* & 0.0129$\pm$0.0010 & 0.972$\pm$0.014 & 0.0268$\pm$0.0005 & 2.87$\pm$0.09 \\
               HD249845 & 72* & 0.0107$\pm$0.0009 & 0.920$\pm$0.012 & 0.0266$\pm$0.0005 & 2.49$\pm$0.07 \\
               HD249845 & 73* & 0.0120$\pm$0.0009 & 0.879$\pm$0.012 & 0.0249$\pm$0.0004 & 2.86$\pm$0.09 \\
               HD252214 &  6  & 0.0189$\pm$0.0029 & 0.239$\pm$0.013 & 0.0757$\pm$0.0057 & 2.03$\pm$0.22 \\
               HD252214 & 33  & 0.0128$\pm$0.0006 & 0.247$\pm$0.007 & 0.0852$\pm$0.0031 & 2.62$\pm$0.15 \\
               HD252214 & 43* & 0.0066$\pm$0.0001 & 0.117$\pm$0.002 & 0.0614$\pm$0.0011 & 3.80$\pm$0.18 \\
               HD252214 & 44* & 0.0067$\pm$0.0001 & 0.102$\pm$0.002 & 0.0564$\pm$0.0010 & 3.94$\pm$0.19 \\
               HD252214 & 71* & 0.0069$\pm$0.0001 & 0.119$\pm$0.002 & 0.0582$\pm$0.0011 & 3.24$\pm$0.13 \\
               HD252214 & 72* & 0.0065$\pm$0.0001 & 0.123$\pm$0.002 & 0.0588$\pm$0.0011 & 3.43$\pm$0.15 \\
               HD254042 & 71  & 0.0116$\pm$0.0005 & 0.399$\pm$0.006 & 0.0396$\pm$0.0008 & 2.39$\pm$0.07 \\
               HD254042 & 72  & 0.0094$\pm$0.0005 & 0.378$\pm$0.005 & 0.0352$\pm$0.0006 & 2.76$\pm$0.09 \\
               HD259865 &  6  & 0.0243$\pm$0.0033 & 0.204$\pm$0.013 & 0.0544$\pm$0.0039 & 2.92$\pm$0.46 \\
               HD259865 & 33  & 0.0139$\pm$0.0008 & 0.205$\pm$0.007 & 0.0660$\pm$0.0034 & 2.24$\pm$0.16 \\
                HD46994 &  6* & 0.0212$\pm$0.0010 & 0.967$\pm$0.012 & 0.0275$\pm$0.0004 & 3.77$\pm$0.14 \\
                HD46994 &  7* & 0.0117$\pm$0.0010 & 0.999$\pm$0.011 & 0.0278$\pm$0.0003 & 4.08$\pm$0.15 \\
                HD46994 & 33* & 0.0076$\pm$0.0007 & 0.563$\pm$0.008 & 0.0232$\pm$0.0004 & 3.15$\pm$0.12 \\
                HD46883 &  6  & 0.0185$\pm$0.0059 & 0.405$\pm$0.021 & 0.0489$\pm$0.0026 & 3.91$\pm$0.62 \\
                HD46883 & 33  & 0.0263$\pm$0.0020 & 0.879$\pm$0.018 & 0.0590$\pm$0.0016 & 2.32$\pm$0.09 \\
                HD47360 &  6  & 0.3619$\pm$0.0256 & 3.231$\pm$0.256 & 0.1111$\pm$0.0112 & 2.06$\pm$0.23 \\
                HD47360 & 33  & 0.0547$\pm$0.0050 & 3.007$\pm$0.110 & 0.1290$\pm$0.0066 & 1.89$\pm$0.08 \\
                HD52463 &  6  & 0.0174$\pm$0.0070 & 0.165$\pm$0.020 & 0.0590$\pm$0.0096 & 1.40$\pm$0.29 \\
                HD52463 &  7  &-0.0115$\pm$0.0171 & 0.185$\pm$0.034 & 0.0410$\pm$0.0079 & 1.09$\pm$0.31 \\
                HD52463 & 33  & 0.0112$\pm$0.0007 & 0.106$\pm$0.005 & 0.0588$\pm$0.0047 & 1.63$\pm$0.13 \\
                HD52463 & 34  & 0.0147$\pm$0.0006 & 0.086$\pm$0.004 & 0.0431$\pm$0.0027 & 2.35$\pm$0.23 \\
                HD53755 &  7  & 0.0353$\pm$0.0052 & 0.317$\pm$0.014 & 0.0341$\pm$0.0014 & 5.05$\pm$0.84 \\
                HD53755 & 33* & 0.0053$\pm$0.0003 & 0.436$\pm$0.004 & 0.0342$\pm$0.0004 & 2.98$\pm$0.07 \\
                HD56876 &  7* & 0.0036$\pm$0.0001 & 0.070$\pm$0.001 & 0.0356$\pm$0.0009 & 2.77$\pm$0.12 \\
                HD56876 & 33* & 0.0030$\pm$0.0001 & 0.061$\pm$0.001 & 0.0294$\pm$0.0006 & 3.36$\pm$0.16 \\
                HD56876 & 34* & 0.0036$\pm$0.0001 & 0.063$\pm$0.001 & 0.0307$\pm$0.0006 & 3.43$\pm$0.16 \\
                HD56876 & 61* & 0.0033$\pm$0.0001 & 0.074$\pm$0.001 & 0.0317$\pm$0.0006 & 3.08$\pm$0.12 \\
                 ALS864 &  7  & 0.0278$\pm$0.0009 & 0.084$\pm$0.006 & 0.1351$\pm$0.0119 & 3.36$\pm$0.70 \\
                 ALS864 &  8  & 0.0381$\pm$0.0017 & 0.169$\pm$0.010 & 0.1521$\pm$0.0030 & 37.5$\pm$23.9 \\
                 ALS864 & 34  & 0.0297$\pm$0.0007 & 0.232$\pm$0.037 & 0.6118$\pm$0.1861 & 1.13$\pm$0.11 \\
                 ALS864 & 61  & 0.0276$\pm$0.0003 & 0.269$\pm$0.044 & 0.4931$\pm$0.1317 & 1.22$\pm$0.10 \\
                HD67536 &  1  & 0.1040$\pm$0.0094 & 0.657$\pm$0.027 & 0.0353$\pm$0.0010 & 9.09$\pm$1.90 \\
                HD67536 &  4* & 0.0163$\pm$0.0006 & 0.703$\pm$0.009 & 0.0324$\pm$0.0005 & 3.28$\pm$0.11 \\
                HD67536 &  7* & 0.0108$\pm$0.0006 & 0.752$\pm$0.010 & 0.0345$\pm$0.0006 & 2.43$\pm$0.06 \\
                HD67536 &  8* & 0.0161$\pm$0.0007 & 0.869$\pm$0.010 & 0.0322$\pm$0.0004 & 3.60$\pm$0.12 \\
                HD67536 &  9* & 0.0119$\pm$0.0006 & 0.849$\pm$0.010 & 0.0334$\pm$0.0005 & 3.10$\pm$0.09 \\
                HD67536 & 10* & 0.0115$\pm$0.0006 & 0.710$\pm$0.009 & 0.0322$\pm$0.0005 & 3.09$\pm$0.10 \\
                HD67536 & 11  & 0.1552$\pm$0.0113 & 0.629$\pm$0.033 & 0.0330$\pm$0.0013 & 5.79$\pm$1.14 \\
                HD67536 & 28* & 0.0094$\pm$0.0006 & 1.016$\pm$0.010 & 0.0332$\pm$0.0003 & 3.69$\pm$0.10 \\
                HD67536 & 31* & 0.0116$\pm$0.0006 & 0.772$\pm$0.008 & 0.0300$\pm$0.0004 & 3.20$\pm$0.09 \\
                HD67536 & 34* & 0.0121$\pm$0.0005 & 0.767$\pm$0.008 & 0.0335$\pm$0.0004 & 3.60$\pm$0.10 \\
                HD67536 & 35* & 0.0066$\pm$0.0006 & 0.855$\pm$0.011 & 0.0377$\pm$0.0007 & 2.29$\pm$0.05 \\
                HD67536 & 37* & 0.0060$\pm$0.0006 & 0.858$\pm$0.011 & 0.0397$\pm$0.0007 & 2.32$\pm$0.06 \\
                HD67536 & 38* & 0.0112$\pm$0.0005 & 0.641$\pm$0.007 & 0.0306$\pm$0.0004 & 2.98$\pm$0.08 \\
                HD67536 & 61* & 0.0102$\pm$0.0006 & 0.630$\pm$0.009 & 0.0322$\pm$0.0005 & 2.97$\pm$0.10 \\
                HD67536 & 62* & 0.0112$\pm$0.0006 & 1.020$\pm$0.009 & 0.0314$\pm$0.0003 & 2.88$\pm$0.06 \\
                HD67536 & 63* & 0.0189$\pm$0.0005 & 0.612$\pm$0.007 & 0.0299$\pm$0.0004 & 3.94$\pm$0.14 \\
                HD67536 & 64* & 0.0129$\pm$0.0005 & 0.598$\pm$0.007 & 0.0323$\pm$0.0005 & 3.25$\pm$0.10 \\
                HD67536 & 68* & 0.0079$\pm$0.0006 & 0.676$\pm$0.008 & 0.0324$\pm$0.0005 & 2.91$\pm$0.09 \\
                HD67536 & 69* & 0.0120$\pm$0.0006 & 0.712$\pm$0.009 & 0.0313$\pm$0.0005 & 2.79$\pm$0.08 \\
    \hline
  \end{tabular}
\end{table*}
\setcounter{table}{0}                
\begin{table*}
  \scriptsize
  \caption{Continued}
  \begin{tabular}{llcccc}
    \hline
                Name & Sector & $C$(mmag)         & $A_0$(mmag)     & $\tau$(d)         & $\gamma$        \\
    \hline
                HD68217 &  7* & 0.0051$\pm$0.0002 & 0.879$\pm$0.015 & 0.2007$\pm$0.0053 & 1.73$\pm$0.03 \\
                HD68217 &  8* & 0.0046$\pm$0.0003 & 0.984$\pm$0.013 & 0.1422$\pm$0.0027 & 2.01$\pm$0.04 \\
                HD68217 & 34* & 0.0131$\pm$0.0003 & 1.032$\pm$0.021 & 0.1486$\pm$0.0046 & 1.57$\pm$0.03 \\
                HD68217 & 35* & 0.0105$\pm$0.0004 & 1.096$\pm$0.028 & 0.1153$\pm$0.0046 & 1.47$\pm$0.03 \\
                HD68217 & 61* & 0.0054$\pm$0.0003 & 0.800$\pm$0.014 & 0.1346$\pm$0.0037 & 1.60$\pm$0.03 \\
                HD68217 & 62* & 0.0030$\pm$0.0002 & 0.738$\pm$0.013 & 0.1662$\pm$0.0046 & 1.72$\pm$0.03 \\
                HD68324 &  7* & 0.0032$\pm$0.0004 & 0.453$\pm$0.010 & 0.0808$\pm$0.0036 & 1.35$\pm$0.03 \\
                HD68324 &  8* & 0.0053$\pm$0.0005 & 0.501$\pm$0.013 & 0.0934$\pm$0.0051 & 1.24$\pm$0.03 \\
                HD68324 &  9* & 0.0013$\pm$0.0004 & 0.451$\pm$0.011 & 0.0876$\pm$0.0045 & 1.19$\pm$0.03 \\
                HD68324 & 34* & 0.0022$\pm$0.0003 & 0.701$\pm$0.015 & 0.1520$\pm$0.0067 & 1.15$\pm$0.02 \\
                HD68324 & 35* & 0.0001$\pm$0.0005 & 0.577$\pm$0.014 & 0.0976$\pm$0.0050 & 1.16$\pm$0.03 \\
                HD68324 & 61* & 0.0035$\pm$0.0004 & 0.519$\pm$0.011 & 0.1024$\pm$0.0046 & 1.22$\pm$0.03 \\
                HD68324 & 62* & 0.0024$\pm$0.0003 & 0.439$\pm$0.009 & 0.0779$\pm$0.0033 & 1.29$\pm$0.03 \\
                HD68962 &  7* & 0.0065$\pm$0.0004 & 0.214$\pm$0.004 & 0.0220$\pm$0.0007 & 2.16$\pm$0.08 \\
                HD68962 &  8* & 0.0094$\pm$0.0004 & 0.235$\pm$0.005 & 0.0215$\pm$0.0006 & 2.38$\pm$0.10 \\
                HD68962 & 34* & 0.0066$\pm$0.0003 & 0.188$\pm$0.003 & 0.0186$\pm$0.0004 & 2.92$\pm$0.12 \\
                HD68962 & 35* & 0.0060$\pm$0.0004 & 0.224$\pm$0.004 & 0.0190$\pm$0.0004 & 2.65$\pm$0.10 \\
                HD68962 & 61* & 0.0067$\pm$0.0003 & 0.228$\pm$0.003 & 0.0198$\pm$0.0003 & 3.14$\pm$0.11 \\
                HD75869 &  8  & 0.0151$\pm$0.0057 & 0.400$\pm$0.033 & 0.0911$\pm$0.0100 & 2.26$\pm$0.37 \\
                HD75869 & 35  & 0.0073$\pm$0.0012 & 0.342$\pm$0.013 & 0.0909$\pm$0.0058 & 1.77$\pm$0.11 \\
                HD75869 & 62  & 0.0062$\pm$0.0002 & 0.282$\pm$0.005 & 0.0835$\pm$0.0021 & 2.10$\pm$0.06 \\
                HD78548 &  8  & 0.0064$\pm$0.0034 & 0.233$\pm$0.013 & 0.0651$\pm$0.0049 & 2.05$\pm$0.24 \\
                HD78548 & 10  & 0.0051$\pm$0.0032 & 0.223$\pm$0.013 & 0.0690$\pm$0.0054 & 1.99$\pm$0.23 \\
                HD78548 & 35* & 0.0030$\pm$0.0002 & 0.265$\pm$0.004 & 0.0663$\pm$0.0015 & 1.87$\pm$0.04 \\
                HD78548 & 37* & 0.0026$\pm$0.0001 & 0.198$\pm$0.002 & 0.0516$\pm$0.0009 & 2.31$\pm$0.06 \\
                HD78548 & 62* & 0.0026$\pm$0.0001 & 0.205$\pm$0.003 & 0.0812$\pm$0.0016 & 2.19$\pm$0.05 \\
                HD78548 & 63* & 0.0028$\pm$0.0001 & 0.166$\pm$0.002 & 0.0656$\pm$0.0012 & 2.60$\pm$0.07 \\
                HD81347 &  9  & 0.0091$\pm$0.0013 & 0.111$\pm$0.007 & 0.0774$\pm$0.0046 & 4.22$\pm$0.80 \\
                HD81347 & 35  & 0.0058$\pm$0.0005 & 0.169$\pm$0.005 & 0.0880$\pm$0.0028 & 3.37$\pm$0.24 \\
                HD81347 & 36  & 0.0054$\pm$0.0004 & 0.149$\pm$0.004 & 0.0935$\pm$0.0032 & 3.13$\pm$0.22 \\
                HD81347 & 62  & 0.0033$\pm$0.0001 & 0.124$\pm$0.002 & 0.0823$\pm$0.0014 & 3.82$\pm$0.18 \\
                HD81347 & 63  & 0.0050$\pm$0.0001 & 0.089$\pm$0.002 & 0.0734$\pm$0.0018 & 3.55$\pm$0.21 \\
                HD87015 & 21* & 0.0066$\pm$0.0002 & 0.249$\pm$0.003 & 0.0423$\pm$0.0008 & 2.38$\pm$0.06 \\
                HD87015 & 45* & 0.0035$\pm$0.0001 & 0.140$\pm$0.002 & 0.0332$\pm$0.0004 & 5.24$\pm$0.24 \\
                HD87015 & 46* & 0.0042$\pm$0.0001 & 0.136$\pm$0.001 & 0.0374$\pm$0.0002 & 20.6$\pm$1.94 \\
                HD87015 & 48* & 0.0056$\pm$0.0001 & 0.157$\pm$0.002 & 0.0377$\pm$0.0002 & 19.0$\pm$1.58 \\
                HD87015 & 72* & 0.0036$\pm$0.0002 & 0.161$\pm$0.002 & 0.0336$\pm$0.0004 & 4.20$\pm$0.17 \\
                HD87152 &  9  & 0.0121$\pm$0.0069 & 0.470$\pm$0.039 & 0.1053$\pm$0.0132 & 1.66$\pm$0.23 \\
                HD87152 & 10  & 0.0264$\pm$0.0062 & 0.483$\pm$0.035 & 0.0911$\pm$0.0089 & 2.13$\pm$0.30 \\
                HD87152 & 36* & 0.0072$\pm$0.0003 & 0.668$\pm$0.008 & 0.0934$\pm$0.0018 & 2.10$\pm$0.04 \\
                HD87152 & 37* & 0.0041$\pm$0.0003 & 0.636$\pm$0.009 & 0.0983$\pm$0.0021 & 2.12$\pm$0.05 \\
                HD87152 & 63* & 0.0092$\pm$0.0003 & 0.803$\pm$0.013 & 0.1454$\pm$0.0040 & 1.64$\pm$0.03 \\
                HD93501 & 10  & 0.0971$\pm$0.0178 & 2.714$\pm$0.104 & 0.0875$\pm$0.0038 & 3.28$\pm$0.32 \\
                HD93501 & 11  & 0.1508$\pm$0.0152 & 1.625$\pm$0.058 & 0.0744$\pm$0.0007 & 63.4$\pm$33.7 \\
                HD93501 & 36  & 0.0734$\pm$0.0048 & 1.812$\pm$0.031 & 0.0716$\pm$0.0005 & 23.9$\pm$3.80 \\
                HD93501 & 37  & 0.0665$\pm$0.0056 & 1.770$\pm$0.051 & 0.0634$\pm$0.0022 & 3.00$\pm$0.20 \\
                HD93501 & 63  & 0.0183$\pm$0.0014 & 1.371$\pm$0.028 & 0.0808$\pm$0.0025 & 2.03$\pm$0.07 \\
                HD93501 & 64  & 0.0339$\pm$0.0017 & 1.817$\pm$0.029 & 0.0603$\pm$0.0015 & 2.09$\pm$0.06 \\
                HD97499 & 10  & 0.0317$\pm$0.0022 & 0.335$\pm$0.012 & 0.1127$\pm$0.0031 & 7.43$\pm$1.25 \\
                HD97499 & 11  & 0.0232$\pm$0.0020 & 0.264$\pm$0.012 & 0.1030$\pm$0.0051 & 3.51$\pm$0.45 \\
                HD97499 & 37  & 0.0148$\pm$0.0008 & 0.308$\pm$0.008 & 0.1034$\pm$0.0030 & 3.73$\pm$0.28 \\
                HD97499 & 64  & 0.0122$\pm$0.0002 & 0.249$\pm$0.004 & 0.1015$\pm$0.0016 & 3.83$\pm$0.16 \\
                HD97913 & 10  & 0.0340$\pm$0.0083 & 0.496$\pm$0.028 & 0.0483$\pm$0.0031 & 2.78$\pm$0.39 \\
                HD97913 & 11  & 0.0195$\pm$0.0108 & 0.640$\pm$0.042 & 0.0698$\pm$0.0064 & 1.78$\pm$0.22 \\
                HD97913 & 37  & 0.0191$\pm$0.0020 & 0.476$\pm$0.014 & 0.0481$\pm$0.0019 & 2.65$\pm$0.18 \\
                HD97913 & 64* & 0.0104$\pm$0.0003 & 0.491$\pm$0.005 & 0.0391$\pm$0.0005 & 2.97$\pm$0.07 \\
               HD108257 & 10* & 0.0045$\pm$0.0001 & 0.082$\pm$0.001 & 0.0257$\pm$0.0004 & 4.64$\pm$0.26 \\
               HD108257 & 11* & 0.0047$\pm$0.0001 & 0.096$\pm$0.001 & 0.0230$\pm$0.0004 & 4.33$\pm$0.23 \\
               HD108257 & 37* & 0.0036$\pm$0.0001 & 0.077$\pm$0.001 & 0.0243$\pm$0.0002 & 8.97$\pm$0.64 \\
               HD125238 & 11* & 0.0024$\pm$0.0002 & 0.313$\pm$0.005 & 0.0845$\pm$0.0020 & 1.95$\pm$0.05 \\
               HD125238 & 38* & 0.0015$\pm$0.0001 & 0.220$\pm$0.004 & 0.1010$\pm$0.0033 & 1.51$\pm$0.03 \\
               HD125238 & 65* & 0.0014$\pm$0.0001 & 0.217$\pm$0.003 & 0.0779$\pm$0.0018 & 2.09$\pm$0.05 \\
               HD133385 & 12  & 0.0138$\pm$0.0040 & 0.317$\pm$0.016 & 0.0654$\pm$0.0043 & 2.18$\pm$0.23 \\
               HD133385 & 38  & 0.0043$\pm$0.0010 & 0.384$\pm$0.009 & 0.0700$\pm$0.0026 & 2.12$\pm$0.10 \\
               HD133385 & 39  & 0.0063$\pm$0.0009 & 0.350$\pm$0.007 & 0.0621$\pm$0.0017 & 2.58$\pm$0.12 \\
               HD133385 & 65  & 0.0054$\pm$0.0003 & 0.357$\pm$0.005 & 0.0609$\pm$0.0012 & 2.35$\pm$0.06 \\
               HD133385 & 66  & 0.0061$\pm$0.0003 & 0.419$\pm$0.005 & 0.0665$\pm$0.0012 & 2.46$\pm$0.06 \\
               HD136298 & 11* & 0.0087$\pm$0.0003 & 0.377$\pm$0.008 & 0.0802$\pm$0.0028 & 1.64$\pm$0.04 \\
               HD136298 & 38* & 0.0025$\pm$0.0002 & 0.370$\pm$0.006 & 0.1184$\pm$0.0040 & 1.31$\pm$0.02 \\
               HD136298 & 65* & 0.0009$\pm$0.0002 & 0.328$\pm$0.007 & 0.1125$\pm$0.0046 & 1.24$\pm$0.02 \\
               HD138485 &Kep. & 0.0302$\pm$0.0008 & 0.035$\pm$0.003 & 0.0487$\pm$0.0039 & 3.72$\pm$0.86 \\
               HD142378 &Kep. & 0.0643$\pm$0.0534 & 0.203$\pm$0.187 & 0.2062$\pm$0.4356 & 0.53$\pm$0.55 \\
               HD143118 & 12* & 0.0028$\pm$0.0002 & 0.549$\pm$0.014 & 0.3872$\pm$0.0203 & 1.12$\pm$0.02 \\
               HD143118 & 65* & 0.0013$\pm$0.0002 & 0.459$\pm$0.008 & 0.1605$\pm$0.0056 & 1.25$\pm$0.02 \\
            CPD-50 9216 & 12  & 0.1145$\pm$0.0193 & 1.146$\pm$0.111 & 0.0829$\pm$0.0112 & 1.63$\pm$0.23 \\
            CPD-50 9216 & 39  & 0.0368$\pm$0.0027 & 0.694$\pm$0.020 & 0.0472$\pm$0.0016 & 3.11$\pm$0.22 \\
            CPD-50 9216 & 66  & 0.0298$\pm$0.0009 & 1.013$\pm$0.018 & 0.0557$\pm$0.0015 & 2.04$\pm$0.06 \\
            CPD-48 8710 & 12  & 0.0670$\pm$0.0054 & 0.379$\pm$0.047 & 0.1759$\pm$0.0364 & 1.46$\pm$0.24 \\
            CPD-48 8710 & 39  & 0.0228$\pm$0.0017 & 0.569$\pm$0.032 & 0.1484$\pm$0.0143 & 1.41$\pm$0.08 \\
            CPD-48 8710 & 66  & 0.0227$\pm$0.0004 & 0.501$\pm$0.012 & 0.0859$\pm$0.0029 & 1.94$\pm$0.06 \\
               HD150745 & 12* & 0.0074$\pm$0.0002 & 0.390$\pm$0.005 & 0.1073$\pm$0.0020 & 2.23$\pm$0.05 \\
               HD150745 & 39* & 0.0036$\pm$0.0001 & 0.274$\pm$0.003 & 0.0915$\pm$0.0016 & 2.13$\pm$0.04 \\
               HD150745 & 66* & 0.0028$\pm$0.0002 & 0.468$\pm$0.006 & 0.1434$\pm$0.0034 & 1.70$\pm$0.03 \\
    \hline
  \end{tabular}
\end{table*}
\setcounter{table}{0}                
\begin{table*}
  \scriptsize
  \caption{Continued}
  \begin{tabular}{llcccc}
    \hline
                Name & Sector & $C$(mmag)         & $A_0$(mmag)     & $\tau$(d)         & $\gamma$        \\
    \hline
               HD164900 & 40* & 0.0061$\pm$0.0003 & 0.433$\pm$0.005 & 0.0744$\pm$0.0008 & 4.69$\pm$0.19 \\
               HD164900 & 53* & 0.0053$\pm$0.0003 & 0.724$\pm$0.005 & 0.0748$\pm$0.0004 & 7.66$\pm$0.27 \\
               HD166197 & 13  & 0.0004$\pm$0.0034 & 0.306$\pm$0.027 & 0.2312$\pm$0.0408 & 1.07$\pm$0.11 \\
               HD166197 & 66* & 0.0030$\pm$0.0001 & 0.164$\pm$0.002 & 0.0568$\pm$0.0007 & 3.16$\pm$0.09 \\
               HD168905 & 13* & 0.0029$\pm$0.0001 & 0.085$\pm$0.001 & 0.0402$\pm$0.0007 & 3.06$\pm$0.11 \\
               HD168905 & 66* & 0.0027$\pm$0.0001 & 0.150$\pm$0.002 & 0.0509$\pm$0.0010 & 2.34$\pm$0.07 \\
               HD180968 & 14* & 0.0123$\pm$0.0006 & 1.442$\pm$0.021 & 0.1202$\pm$0.0031 & 1.58$\pm$0.03 \\
               HD180968 & 40* & 0.0177$\pm$0.0006 & 1.093$\pm$0.012 & 0.0574$\pm$0.0008 & 2.94$\pm$0.08 \\
               HD180968 & 54* & 0.0053$\pm$0.0008 & 1.374$\pm$0.024 & 0.0855$\pm$0.0026 & 1.60$\pm$0.03 \\
               HD192968 & 14  & 0.0127$\pm$0.0034 & 0.315$\pm$0.056 & 0.5142$\pm$0.1992 & 0.94$\pm$0.14 \\
               HD192968 & 15  & 0.0198$\pm$0.0024 & 13.21$\pm$176.7 & 216.06$\pm$3678.3 & 0.84$\pm$0.15 \\
               HD192968 & 41* & 0.0052$\pm$0.0001 & 0.138$\pm$0.004 & 0.0944$\pm$0.0049 & 1.38$\pm$0.04 \\
               HD192968 & 55* & 0.0060$\pm$0.0001 & 0.012$\pm$0.001 & 0.0098$\pm$0.0004 & 5.16$\pm$0.77 \\
 Cl* Berkeley 86 HG 261 & 14  & 0.0395$\pm$0.0046 & 0.729$\pm$0.077 & 0.3663$\pm$0.0713 & 1.19$\pm$0.11 \\
 Cl* Berkeley 86 HG 261 & 15  & 0.0806$\pm$0.0030 & 0.411$\pm$0.036 & 0.2415$\pm$0.0301 & 2.03$\pm$0.28 \\
 Cl* Berkeley 86 HG 261 & 41  & 0.0314$\pm$0.0007 & 0.480$\pm$0.029 & 0.4920$\pm$0.0545 & 1.38$\pm$0.08 \\
 Cl* Berkeley 86 HG 261 & 55  & 0.0358$\pm$0.0010 & 0.437$\pm$0.030 & 0.3809$\pm$0.0500 & 1.24$\pm$0.08 \\
               HD193794 & 14  & 0.0265$\pm$0.0069 & 0.759$\pm$0.032 & 0.0747$\pm$0.0042 & 2.23$\pm$0.20 \\
               HD193794 & 15  & 0.0393$\pm$0.0061 & 0.649$\pm$0.026 & 0.0623$\pm$0.0029 & 2.88$\pm$0.29 \\
               HD193794 & 41* & 0.0070$\pm$0.0003 & 0.716$\pm$0.006 & 0.0595$\pm$0.0007 & 2.52$\pm$0.05 \\
               HD193794 & 55* & 0.0071$\pm$0.0003 & 0.584$\pm$0.005 & 0.0519$\pm$0.0006 & 2.69$\pm$0.05 \\
               HD198781 & 15* & 0.0118$\pm$0.0005 & 0.733$\pm$0.008 & 0.0452$\pm$0.0005 & 3.58$\pm$0.11 \\
               HD198781 & 17* & 0.0131$\pm$0.0005 & 0.642$\pm$0.007 & 0.0437$\pm$0.0005 & 3.90$\pm$0.13 \\
               HD198781 & 18* & 0.0108$\pm$0.0004 & 0.625$\pm$0.007 & 0.0443$\pm$0.0006 & 3.59$\pm$0.11 \\
               HD198781 & 24* & 0.0110$\pm$0.0004 & 0.624$\pm$0.006 & 0.0457$\pm$0.0005 & 4.04$\pm$0.12 \\
               HD198781 & 56* & 0.0042$\pm$0.0004 & 0.554$\pm$0.005 & 0.0428$\pm$0.0005 & 3.82$\pm$0.11 \\
               HD198781 & 57* & 0.0056$\pm$0.0004 & 0.642$\pm$0.007 & 0.0439$\pm$0.0005 & 3.42$\pm$0.10 \\
               HD198781 & 58* & 0.0044$\pm$0.0004 & 0.592$\pm$0.006 & 0.0480$\pm$0.0007 & 2.79$\pm$0.07 \\
               HD201819 & 15* & 0.0094$\pm$0.0003 & 0.419$\pm$0.004 & 0.0325$\pm$0.0003 & 3.87$\pm$0.11 \\
               HD201819 & 55* & 0.0045$\pm$0.0003 & 0.436$\pm$0.004 & 0.0378$\pm$0.0005 & 2.77$\pm$0.06 \\
               HD201819 & 56* & 0.0032$\pm$0.0002 & 0.373$\pm$0.003 & 0.0358$\pm$0.0004 & 2.93$\pm$0.06 \\
               HD207308 & 16  &-0.0004$\pm$0.0069 & 0.219$\pm$0.014 & 0.0304$\pm$0.0020 & 2.23$\pm$0.32 \\
               HD207308 & 17  &-0.0091$\pm$0.0107 & 0.245$\pm$0.021 & 0.0346$\pm$0.0031 & 1.64$\pm$0.27 \\
               HD207308 & 24  & 0.0036$\pm$0.0045 & 0.184$\pm$0.009 & 0.0282$\pm$0.0014 & 2.81$\pm$0.36 \\
               HD207308 & 56  & 0.0048$\pm$0.0002 & 0.152$\pm$0.002 & 0.0311$\pm$0.0007 & 2.27$\pm$0.07 \\
               HD207308 & 57  & 0.0042$\pm$0.0003 & 0.185$\pm$0.003 & 0.0342$\pm$0.0009 & 2.11$\pm$0.07 \\
               HD207308 & 58  & 0.0037$\pm$0.0002 & 0.167$\pm$0.003 & 0.0331$\pm$0.0008 & 2.10$\pm$0.06 \\
          LS III +57 89 & 16  & 0.0394$\pm$0.0023 & 0.413$\pm$0.052 & 0.5356$\pm$0.1214 & 1.33$\pm$0.16 \\
          LS III +57 89 & 17  & 0.0433$\pm$0.0018 & 0.429$\pm$0.027 & 0.2379$\pm$0.0181 & 2.41$\pm$0.24 \\
               HD216092 & 16  & 0.0152$\pm$0.0150 & 0.341$\pm$0.029 & 0.0280$\pm$0.0023 & 2.39$\pm$0.46 \\
               HD216092 & 17  & 0.0093$\pm$0.0258 & 0.465$\pm$0.051 & 0.0311$\pm$0.0032 & 1.75$\pm$0.36 \\
               HD216092 & 56  & 0.0076$\pm$0.0006 & 0.460$\pm$0.007 & 0.0374$\pm$0.0010 & 1.96$\pm$0.05 \\
               HD216092 & 57  & 0.0120$\pm$0.0006 & 0.481$\pm$0.007 & 0.0316$\pm$0.0007 & 2.32$\pm$0.07 \\
           NGC 7654 485 & 17  & 0.0679$\pm$0.0050 & 0.606$\pm$0.057 & 0.2505$\pm$0.0403 & 1.45$\pm$0.17 \\
           NGC 7654 485 & 18  & 0.0603$\pm$0.0046 & 0.606$\pm$0.060 & 0.2399$\pm$0.0377 & 1.61$\pm$0.20 \\
           NGC 7654 485 & 24  & 0.0352$\pm$0.0025 & 0.251$\pm$0.022 & 0.1780$\pm$0.0219 & 2.28$\pm$0.39 \\
           NGC 7654 485 & 57  & 0.0434$\pm$0.0004 & 0.341$\pm$0.011 & 0.1715$\pm$0.0083 & 2.28$\pm$0.14 \\
           NGC 7654 485 & 58  & 0.0577$\pm$0.0005 & 0.626$\pm$0.033 & 0.2598$\pm$0.0194 & 1.80$\pm$0.10 \\
               HD223145 &  1* & 0.0051$\pm$0.0001 & 0.295$\pm$0.002 & 0.1008$\pm$0.0010 & 4.01$\pm$0.11 \\
               HD223145 &  2* & 0.0045$\pm$0.0001 & 0.268$\pm$0.002 & 0.0955$\pm$0.0004 & 12.5$\pm$0.61 \\
               HD223145 & 28* & 0.0035$\pm$0.0001 & 0.330$\pm$0.004 & 0.0942$\pm$0.0013 & 3.33$\pm$0.09 \\
               HD223145 & 29* & 0.0058$\pm$0.0001 & 0.302$\pm$0.003 & 0.0922$\pm$0.0012 & 3.46$\pm$0.10 \\
               HD223145 & 68* & 0.0038$\pm$0.0001 & 0.309$\pm$0.004 & 0.0824$\pm$0.0017 & 2.10$\pm$0.05 \\
               HD223145 & 69* & 0.0032$\pm$0.0001 & 0.329$\pm$0.005 & 0.0927$\pm$0.0024 & 1.77$\pm$0.04 \\
    \hline
  \end{tabular}
\end{table*}

\section{Properties of isolated signals}

  \begin{sidewaystable*}
  \scriptsize
  \caption{Properties of isolated signals detected in our sample.
 \label{domf}}
  \begin{tabular}{lllll}
    \hline
Name    & \multicolumn{2}{c}{Low Frequencies (0.5--6.0\,d$^{-1}$) in d$^{-1}$ (amplitudes in mmag)} & High frequencies (6.--15.\,d$^{-1}$) & Very high frequencies ($>15$\,d$^{-1}$) \\
        & dominant peak & other prominent peaks  & frequencies in d$^{-1}$ (amplitudes in mmag) & frequencies in d$^{-1}$ (amplitudes in mmag) \\
\hline
HD5882	           & 2.540 (2.5--2.9)  & 4.86 (0.3--0.5) & 9.874 (0.2--0.6), 14.63 (0.1)	&    \\
NGC 869 133	   &   	               &   	         &    	                                & 15.060 (0.3--0.5) \\
BD+56 538	   & 5.516 (0.8--0.9)  & 0.960 (1.9) in Sector 58 & 7.452 (0.2) in Sector 58    &    \\
HD14250	           & 0.812 (24.3--25.5)&   	         &   	                                &    \\
BD+62 657	   &   	               & 5.034 (0.2)	 & 8.692 (0.3--0.4), 9.152 (0.2), 9.692 (0.2), 12.984 (0.1--0.2), 14.520 (0.5--0.6) & 26.572(0.12) in Sector 59 \\
HD34748	           &   	               &                 & 7.618 (0.2--0.3), 9.310 (0.1--0.2)   &    \\
HD35532	           & 1.764 (2.1--4.0) + 3.524 (1.9--4.7)& 3.090 (1.2--1.5) except Sector 43     &    \\
HD35777	           & 1.716 (2.5) + 3.428 (2.5)           &   	              & 6.856 (0.2), 8.132 (0.2), 9.888 (0.2), 12.000 (0.09)$^{a1}$ & 18.128 (0.05)$^{a2}$ \\
HD37303	           & 2.140 (0.35)      & 1.620 (0.1), 1.870 (0.2), 4.252 (0.1) & 11.788 (0.3), 13.430 (0.2), 14.980 (0.07) & 19.292(0.3), 23.632(0.05)$^b$ \\
HD37397	           & 1.744 (9.0)       & 3.492 (5.4)     & 6.982 (0.5), 10.468 (0.2), 12.210 (0.2) & 15.708 (0.13)\\
HD37674	           & 2.764 (5.8)       & 3.304 (1.1)     & 6.076 (2.1), 6.612 (0.6), 12.144 (0.4), 12.664 (0.5) &    \\
HD252214	   & 1.456 (1.6--1.9)  & 0.728 (0.7--1.0), 2.904 (0.2--0.3), 3.032 (0.2--0.3)&  &    \\
HD254042	   &                   & 4.990 (0.8--1.0)&    	                                &    \\
HD259865	   &                   & 0.582 (1.0--1.4)& 7.640 (0.1--0.2)                     &    \\
HD46883	           & 1.764 (1.7--2.3) + 2.660 (1.4--1.5)&                  &                    &    \\
HD52463	           &   	               & 2.354(0.4) + 3.286 (0.3--0.4) in Sectors 6--7 &        &    \\
HD53755	           & 1.536 (4.5--5.9) + 3.388 (5.1--5.9)& 2.86 (1.4--1.8), 3.76 (1.1--3.1) &    &    \\
HD56876	           & 1.724 (1.1)       &  3.448 (0.3), 4.552 (0.3--0.4)     & 6.288 (0.2), 8.806 (0.17) & 22.400 (0.02), 24.124 (0.02--0.03) \\
ALS 864	           & 0.835 (0.4--0.5)  &   	         &   	                                &    \\
HD68324	           &                   &   	         & 8.778 (1.5--2), 10.704 (3.0--3.3)	&    \\
HD87015	           & 3.532 (2.7--3.1)  & 1.970 (0.6--1.2)&                                      &    \\
HD87152	           & 1.086 (2.2--5.5) + 1.290 (1.9--3.9)& 2.536 (1.2--1.8) &                    & 37.184 (0.02--0.05) \\
HD97499	           & 1.046 (1.8--2.0)  & 0.750 (1.4--1.7), 3.010 (0.2--0.3)&                    &    \\
HD97913	           & 2.176 (4.0--5.9)  & 5.534 (1.4--1.9)& 11.072 (0.2) in Sector 64	        &    \\
HD108257	   &                   &                 &                                      & 28.120 (0.03) in Sector 10\\
HD133385	   & 1.080 (1.1--3.8)  & 2.000 (1.0--1.5)&                                      &    \\
HD136298	   & 5.900 (1.2--1.4)  & 1.276 (0.9--1.6), 2.538 (0.6--0.8), 2.700 (0.4--0.5), $^c$ & 8.096 (0.2--0.3), 8.580 (0.02), 9.300 (0.2), 9.770 (0.1--0.2), 10.576 (0.1--0.2), 11.746 (0.2) &    \\
HD138485	   & 1.943 (1.7)       & 0.560 (0.7), 1.607 (0.4), 1.655 (0.4), $^d$ & 6.155 (0.2), 11.401 (0.15) &    \\
HD142378	   &                   & 1.451 (1.0), 1.730 (0.9), 3.865 (1.3)&                 &    \\
HD143118	   &                   &   	            & 6.060 (0.2--0.3), 6.566 (0.2--0.3)$^{e1}$ & 26.234 (0.03--0.04)$^{e2}$ \\
CPD--50 9216	   &                   & 1.536 (3.6--3.9)&                                      &    \\
CPD--48 8710	   & 1.688 (2.4--2.6)  & 3.376 (0.9--1), 5.060 (0.4--0.5)  & 6.752 (0.4--0.6)   & 18.396 (0.3) \\
HD150745	   &                   & 0.938 (1.1--1.4)&                                      &    \\
HD164900	   & 0.65 (3.5--4.1) + 1.406 (3.4--5.2) & 1.842 (2.2--2.6), 4.610 (0.3)    &    &    \\
HD168905	   &                   & 1.184 (0.8) in Sector 66 & 11.896(0.2) in Sector 66	&    \\
HD180968	   & 0.640 (7.6--9.5)  & 3.246 (1.6--2.8)&                                      &    \\
HD192968	   &                   & 5.964 (0.08--0.1)$^{f1}$ & 9.392 (0.2--0.4), 10.012 (0.2--0.9), 13.504 (0.11--0.13) & 20.364 (0.07--0.13), 20.604 (0.13), $^{f2}$ \\
Cl* Berk. 86 HG 261&	               & 5.926 (0.3--0.4)& 6.926 (0.3), 13.282 (0.2)            &    \\
HD198781           & 1.536 (5.3--7.0)  & 2.554 (4.2--5), 2.914 (1.9--2.2)  &                    &    \\
HD201819	   & $^g$              &                                   &                    &    \\
HD207308	   & 2.822 (0.8--1.5) + 4.510 (0.8--1.0) & 4.304 (0.7--0.9) in Sectors 16 and 17 & 6.86 (0.8--1.), 9.118 (0.2, except Sect. 16, 17)  &    \\
HD223145	   &                   & 5.010 (0.1--0.5)                  &                    &    \\
    \hline
  \end{tabular}
    \tablefoot{
      $^{a1}$ plus 13.716(0.06) in Sector 32;
$^{a2}$ plus 15.432 (0.06), 20.576 (0.02), 25.716 (0.02), and 29.148 (0.02) in Sector 32;
$^b$ plus 26.308 (0.05) and 34.272 (0.07) in Sector 6, and 24.896 (0.05) in Sector 32;
$^c$ 4.710 (0.03), 4.870 (0.02);
$^d$ 1.851 (0.4), 2.076 (0.8), 2.561 (0.5), 3.550 (0.4), 3.685 (0.8), 5.630 (0.3);
$^{e1}$ plus 9.740 (0.15) in Sector 12 and 6.728 (0.2) in Sector 12;
$^{e2}$ not formally significant but coherent between sectors;
$^{f1}$ plus 1.628 (0.5) and 1.868 (0.4) in Sector 41;
$^{f2}$ 29.314 (0.07), 29.632 (0.03--0.06), 33.966 (0.03), 36.074 (0.05), 37.200 (0.03--0.04), 40.190 (0.03--0.05), 40.726 (0.03) plus 32.396 (0.07) in Sector 41 and 35.376 (0.03) in Sector 55;
$^g$ 1.448 (3.4) in Sector 15, 2.766 (1.7--1.9) in Sectors 55 and 56.
    }
\end{sidewaystable*}

  \section{A note on the Be stars 25\,Ori, 25\,Cyg and V434\,Aur}

\begin{figure*}
  \centering
  \includegraphics[width=18cm]{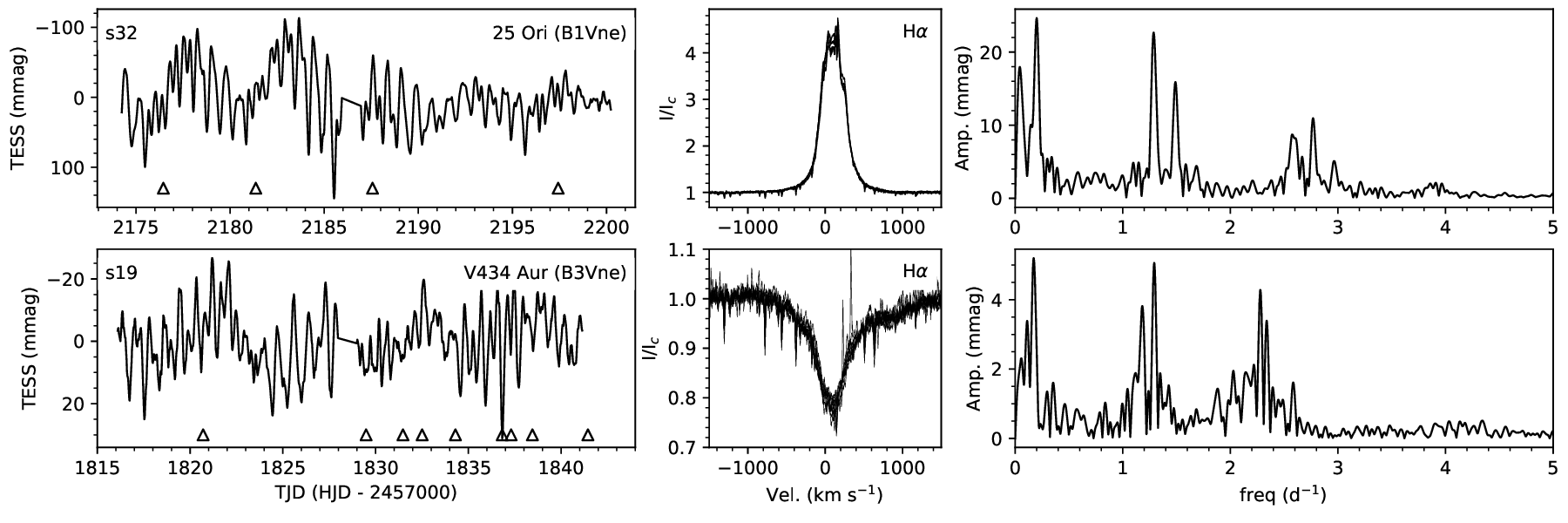}
  \caption{\te\ photometry and contemporaneous spectroscopy for two Be stars, 25\,Ori (top) and V434\,Aur (bottom). The left panels show the \te\ light curve from the indicated sector (at the top left), with triangles corresponding to dates of NRES spectroscopic observations. The middle panels show the spectral region centred on H$\alpha$ for each spectroscopic epoch shown in the left panels. The right panels contain the frequency spectra built from the \te\ photometry. 
  }
  \label{be}
\end{figure*}

In order to assess potential contributions to photometric signals from circumstellar material, spectroscopic observations were taken of select Be stars during their \te\ observing windows. The spectroscopy can then be used to infer whether or not any disk was present during the \te\ observations (see Labadie-Bartz et al., in preparation). Two example Be stars observed in this fashion, 25\,Ori and V434\,Aur, are shown in Fig.~\ref{be}. For 25\,Ori, the spectra show strong H$\alpha$ emission, indicating the presence of a dense disk, while for V434\,Aur there is no emission in H$\alpha$ (nor in any other lines) indicating that no disk was present at the time of \te\ observations. However, the light curves and frequency spectra of these two stars are qualitatively the same, despite 25\,Ori having a strong disk, and V434\,Aur having no disk. 

\begin{figure*}
  \centering
  \includegraphics[width=18cm]{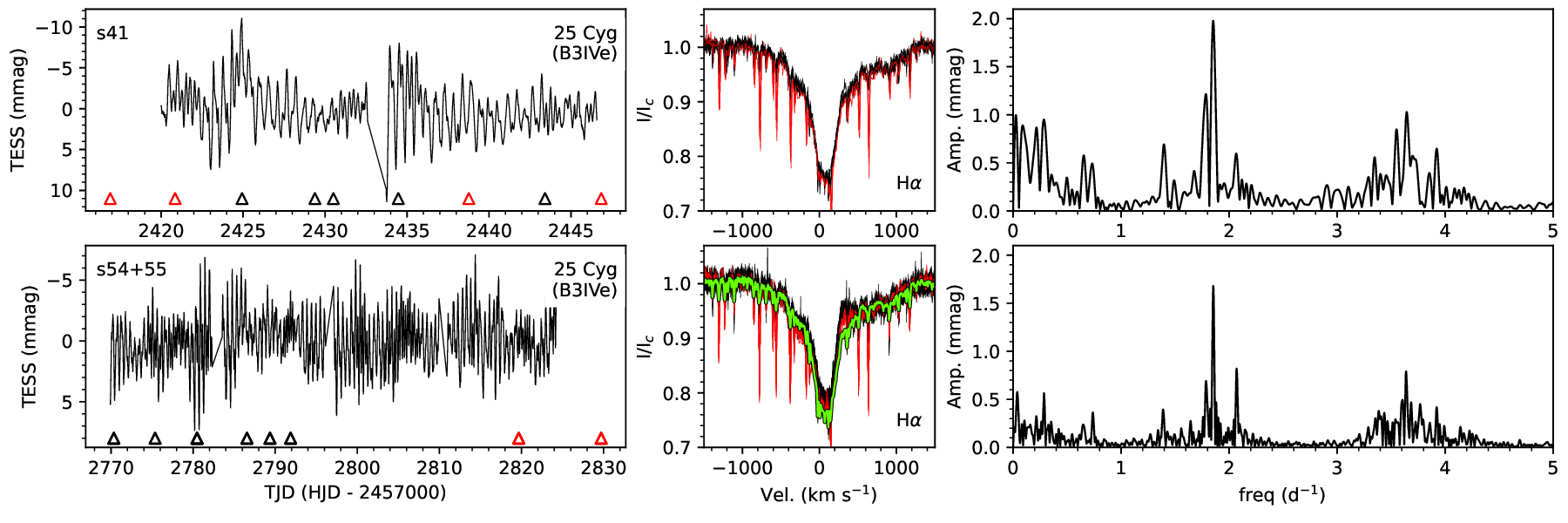}
  \caption{Similar to Fig.~\ref{be}, but comparing two epochs for the same Be star (25\,Cyg). The black (respectively red) triangles in the left panels indicate spectra from NRES (respectively DAO). The same colours are used in the middle panels for data from the two telescopes. The green line in the bottom-middle panel is the average of the NRES H$\alpha$ profile taken during Sector 41 (i.e. during an emission-free phase). 
  }
  \label{be2}
\end{figure*}

A similar comparison can be made for observations of the same star, 25\,Cyg, in two different \te\ sectors (Fig.~\ref{be2}). During Sector 41, the system had no emission (and is thus without any disk material). In Sectors 54 and 55, H$\alpha$ displayed weak emission, indicating a low density (dissipating) disk. The photometric variability at these two epochs is essentially the same, with only mild variations in the strength of certain frequencies, but the overall structure of the FGs remains the same. This suggests that for this star, the presence of a (weak) disk has no major impact on the signals encoded in \te. 

The spectroscopic data used in the above examples were taken with two different instruments. Echelle spectra were obtained with the Network of Robotic Echelle Spectrographs (NRES, ${R\sim53000}$) attached to the 1\,m telescopes of the Las Cumbres Observatory Global Telescope network \citep{2013PASP..125.1031B}, including at the Wise observatory, the South African Astronomical Observatory, and McDonald Observatory. Observations were also obtained from the Dominion Astrophysical Observatory (DAO) 1.2m telescope equipped with the McKeller spectrograph (with resolving power ${R\sim17600}$), which covers H$\alpha$ and \ion{He}{I}\,$\lambda$6678 in the chosen observing mode \citep{2014RMxAC..45...69M}. 

\end{appendix}
\end{document}